\documentclass[aip,graphicx]{revtex4-1}
\usepackage{graphicx}
\usepackage{dcolumn}
\usepackage{bm}
\usepackage{amsmath,scalefnt}
\usepackage{subfigure}
\usepackage{float}
\numberwithin{equation}{section}
\usepackage{natbib}

\begin{document}
\title{Optoelectronic response calculations in the framework of  \textbf{k} $\cdot${} \textbf{p} coupled to Non-equilibrium Green's functions for 1D systems in the ballistic limit}%

\author{Andrei Buin}%
\affiliation{Department of Electrical and Computer Engineering, University of Waterloo, Waterloo, Ontario N2L 3G1, Canada}%

\author{Amit Verma}%
\affiliation{Department of Electrical Engineering and Computer Science, Texas A$\&$M University – Kingsville, Kingsville, Texas 78363, USA}%

\author{Simarjeet Saini}%
\affiliation{Department of Electrical and Computer Engineering, University of Waterloo, Waterloo, Ontario N2L 3G1, Canada}%

\email{phquanta@gmail.com}%


\begin{abstract}
  We present theory of the carrier-optical interaction in 1D systems based on the nonequilibrium Green’s function formalism in the 4x4 \textbf{k} $\cdot${} \textbf{p} model.  As a representative parameters we chose the GaAs. Although theory is presented in 4x4kp many subbands, results and discussion section is based on the simplified model such as 2x2 kp model (two transverse modes). Even though 2x2 kp model is simple enough it shows many phenomena that have not been seen before. We focus mainly on the ballistic extraction of photogenerated free carriers at the radiative limit which is described by  the self-energy term derived in dipole approximation and solved in self-consistent manner with Keldysh quantum kinetic equations. Any relaxation or non-radiative recombination mechanisms as well as excitonic features are neglected. Effect of non-locality of electron-photon self energy term is  considered and discussed. Spontaneous emission is also considered and shown to be small in short devices under medium bias conditions. Electron and hole spatial current oscillations are seen and discussed. It is shown that neglecting off-diagonal correlation in the band index not only produces quantitatively wrong results but it also alters the qualitative picture. All simulations are done in the  full-rank approximation , with all spatial and band correlation effects are kept intact. This allows us to study not only quantitative effects but also qualitative behaviour.
\end{abstract}
\maketitle
\section{Indroduction}

The past several years has seen a growing interest in nanowires (NWs) such as {SiNWs}\cite{Ramanujam,verma}, {GeNWs}\cite{Cao}, and {GaAsNWs} because of their excellent optoelectronic properties \cite{Lieber}. As few examples, recent experimental work \cite{thunich} on the photocurrent response of freely suspended single 140 nm {GaAsNWs} has shown current as high as $\sim$0.45 nA for a titanium:sapphire laser light intensity of 100 $W/cm^{2}$. Experimental work on the effect of strain on {GaAsNWs}, approximately 80 nm in diameter\cite{zardo}, and theoretical work on much smaller diameter {SiNWs}\cite{Dar}, have also shown a direct-to-indirect bandgap transition, which can potentially be used for laser applications. At the same time, it has also been found that surface passivation of the {GaAsNW} with {Al$_{x}$Ga$_{1-x}$As} increases the photoluminescence (PL) lifetime, and minority carrier diffusion lengths, significantly\cite{demichel,Chang}. The bandgap in {GeNWs} is also found to be dependent upon the type of surface passivation as well as strain \cite{C3CP}, which has a consequence on the optoelectronic response of the NW. Concomitantly, GaAs p-i-n NW structures have also shown excellent solar power harvesting capability\cite{colombo}. The above results (as well as several references contained therein) highlight the significance of obtaining a detailed understanding of the photo response of {NWs} and 1D devices.  As these nanostructures are being used for image photo detectors, calculations of the photo current response become important.  Of particular importance is the behavior of smaller diameter {NWs}, in keeping with the trend towards smaller feature sizes. Over the years several theoretical work have been reported to understand the optoelectronic response of {NWs}\cite{AAA,Fedoseyev}. Just to  name a few advanced works in the field of NEGF coupled to photonic field which recently appeared are the works  of Aerberhard et al.\cite{Aerb2,AerbSuper} , Steiger\cite{SteigerPhD}  and Henrickson et al.\cite{henrickson} and Stewart et.al \cite{Stewart}  which use either Tight Binding(TB) or bulk 2D \textbf{k} $\cdot${} \textbf{p} modeling. The limitation of the TB is the system size, whereas the limitation of the bulk 2D \textbf{k} $\cdot${} \textbf{p} system is applicability to 2D systems such as quantum wells, superlattices. In this work we discuss an approach suitable for modeling the photo current response of sub-10 nm diameter {NWs}. The basis of our work is the band structure calculation by utilizing a 1D 4x4 \textbf{k} $\cdot${} \textbf{p} model, with transport calculations utilizing non-equilibrium Green's function (NEGF) formalism. For small structures, semiclassical simulations, such as Monte Carlo, are reasonably accurate, but they may not capture the details of charge distribution in its entirety, particularly in the problem being addressed. On the other hand, NEGF based quantum mechanical approach may provide a more accurate estimation especially in the phase-coherent regime. Moreover, NEGF allows to incorporate phase-breaking(not considered here) processes vis self-energies.  To our knowledge this is the first work which couples 4x4 \textbf{k} $\cdot${} \textbf{p} and NEGF to compute the photo response of the 1D nanostructures. We have used 4x4 \textbf{k} $\cdot${} \textbf{p} (applicable to direct band gap materials) to keep things simple, although conceptually there is no restriction and the model can be easily transferred to the indirect band-gap materials such as Si, Ge by using a larger dimensional \textbf{k} $\cdot${} \textbf{p} such as 15x15, 24x24, 30x30 \textbf{k} $\cdot${} \textbf{p} models for sub-10 nm 1D devices.  This approach takes into account  correlations between different band indices as well as spatial correlation  allowing us to study the effect of non-locality of electron-photon self-energy. We believe that the proposed method provides a good compromise between computational speed and modeling complexity. The paper is divided into different sections. Section II focuses on the theory, particularly band structure calculation, electron-photon interaction, transport formalism, as well as mode-space approach and physical observables. Device setup and numerical parameters are discussed in Section III. Section IV comprises of results and discussion, and conclusions are drawn in Section V.

\section{Theory}
\label{sec:Theory}
\subsection{Hamiltonian}
Starting point of the work was the calculation of the band structure using the original Kane \cite{Kane82} 4x4 \textbf{k} $\cdot${} \textbf{p} scheme and using GaAs as a representative material. For the computation of the photoresponse (discussed below), we use a modified 2x2 scheme (or two subband model).   Originally \textbf{k} $\cdot${} \textbf{p} was done for the direct bandgap materials, although usage of it has been extended to indirect bandgap materials (Si and Ge)\cite{kp714,KpSi,kpSi1,kpSi30,kp24,Cardona}, and one can adapt the present method to originally indirect bulk
materials.

Hamiltonian is given in basis of cell-periodic zone centered (\textbf{k}=0) Bloch functions
$\left\vert u_{j}\right\rangle =\{\left\vert S\uparrow \downarrow
\right\rangle ,\left\vert X\uparrow \downarrow \right\rangle ,\left\vert
Y\uparrow \downarrow \right\rangle ,\left\vert Z\uparrow \downarrow
\right\rangle \}$ \cite{kp8}

\footnotesize
\begin{equation}
\mathbf{H}_{0,bulk}=\\
\begin{pmatrix}
E_{g}+\gamma _{a}(k_{x}^{2}+k_{y}^{2}+k_{z}^{2}) & iPk_{x} & iPk_{y} & iPk_{z} \\
-iPk_{x} & Lk_{x}^{2}+M(k_{y}^{2}+k_{z}^{2}) & Nk_{x}k_{y} & Nk_{x}k_{z} \\
-iPk_{y} & Nk_{x}k_{y} & Lk_{y}^{2}+M(k_{x}^{2}+k_{z}^{2}) & Nk_{y}k_{z} \\
-iPk_{z} & Nk_{x}k_{z} & Nk_{y}k_{z} & Lk_{z}^{2}+M(k_{x}^{2}+k_{y}^{2})%
\end{pmatrix}%
\end{equation}
\normalsize
where $\gamma _{a}=\frac{\hbar^2}{2m_{0}}+F$,  $F$ is the effect of remote bands\cite{Kane82}
, $E_{p}=2m_{0}P^{2}/\hbar^{2}$   and
\begin{equation}
L=-\frac{\hbar^{2}}{2m_{0}}(\gamma _{1}+4\gamma_{2})
\end{equation}
\begin{equation}
M=-\frac{\hbar^{2}}{2m_{0}}(\gamma _{1}-2\gamma _{2})
\end{equation}
\begin{equation}
N=-\frac{\hbar^{2}}{2m_{0}}6\gamma _{3}
\end{equation}
\bigskip
with $\gamma _{1},\gamma _{2},\gamma _{3}$  being modified Luttinger parameters are related to the original Luttinger parameters $ (\gamma _{1}^{L},\gamma _{2}^{L},\gamma _{3}^{L})$ by
\begin{equation}
\gamma_{1}=\gamma _{1}^{L}-\frac{E_{p}}{3E_{G}}
\end{equation}
\begin{equation}
\gamma _{2}=\gamma _{2}^{L}-\frac{E_{p}}{6E_{G}}
\end{equation}
\begin{equation}
\gamma _{3}=\gamma _{3}^{L}-\frac{E_{p}}{6E_{G}}
\end{equation}
\bigskip
where ($E_{p},E_{G},\gamma _{1}^{L},\gamma _{2}^{L},\gamma _{3}^{L}$)\cite{Foreman}
are specific material parameters. Renormalisation is required so as to subtract effects of conduction band in the original 3x3 \textbf{k} $\cdot${} \textbf{p} model \cite{ModifLuttinger}.
Making transformation from $k$-space representation to real space
representation one has to use momentum operators which are given by
\begin{equation}
k_{x}=-i\hbar\frac{\partial }{\partial x},k_{y}=-i\hbar\frac{\partial }{\partial y},
k_{z}=-i\hbar\frac{\partial }{\partial z}
\end{equation}
Since sine waves naturally satisfy infinite barrier boundary conditions, they are chosen as basis functions in  the transverse direction. This corresponds
to the situation of free-standing p-i-n junction.
Along the transport direction ($x$-axis) we adopt the following basis functions
\begin{equation}
\chi _{i}(x)=\frac{1}{\sqrt{\Delta }}(\Theta(x-x_{i})-\Theta (x-x_{i+1}))
\end{equation}
\bigskip
where $\Theta $ is the Heaviside function, and $\Delta $ is the interlayer
spacing. Electronic wavefunction in the aforementioned basis is
written as
\begin{equation}
\label{eq:PSI}
\left\vert \Psi\right\rangle=\sum\limits_{j,p,q,i} a_{j,p,q}(x)\sin (k_{p}y)\sin
(k_{q}z)\chi _{i}(x)\left\vert u_{j}\right\rangle
=\sum\limits_{j,p,q,i}a_{j,p,q}(x)\left\vert j,p,q,i\right\rangle
\end{equation}
where $j=\{1,2,3,4\}$, $p=1..N_{p},$ $q=1...N_{q}$ and $k_{p}=\frac{%
p\pi }{L_{y}},k_{q}=\frac{q\pi }{L_{z}},$ $i=1..N_{x}$ with $p$ denoting $y$ and $q$
denoting $z$. The total Hamiltonian \cite{shinMain} in basis $\left\vert
j,p,q,i\right\rangle $
\begin{equation}
\mathbf{H}_{tot}=\\
\begin{pmatrix}
\mathbf{H}_{1} & \mathbf{W} & 0 & ... & 0 \\
\mathbf{W}^{\dag } & \mathbf{H}_{2} & \mathbf{W} & ... & 0 \\
0 & \mathbf{W}^{\dag } & \mathbf{H}_{3} & \ddots & 0 \\
0 & ... & \ddots & \ddots & \mathbf{W} \\
0 & ... & 0 & \mathbf{W}^{\dag } & \mathbf{H}_{N_{x}}%
\end{pmatrix}
\end{equation}
where $\mathbf{H}_{i}$ \ is the block matrix of the size $4N_{p}N_{q}$ and
given by
\begin{equation}
\mathbf{H}_{i}=\\
\begin{pmatrix}
\mathbf{H}_{\{1,1\},\{1,1\}} & \mathbf{H}_{\{1,1\},\{1,2\}} & ... & ... &
\mathbf{H}_{\{1,1\},\{N_{p},N_{q}\}} \\
\mathbf{H}_{\{1,2\},\{1,1\}} & ... & \mathbf{...} & ... & \mathbf{H}%
_{\{1,2\},\{N_{p},N_{q}\}} \\
... & \mathbf{...} & \mathbf{H}_{\{p,q\},\{p^{\prime },q^{\prime }\}} & ...
& ... \\
\mathbf{...} & ... & ... & ... & \mathbf{...} \\
\mathbf{H}_{\{N_{p},N_{q}\},\{1,1\}} & ... & ... & \mathbf{H}
_{\{N_{p},N_{q}\},\{N_{p},N_{q}-1\}} & \mathbf{H}_{\{N_{p},N_{q}\},%
\{N_{p},N_{q}\}}%
\end{pmatrix}
\end{equation}
with
\footnotesize
\begin{equation}
\mathbf{H}_{(p,q),(p^{\prime },q^{\prime })}=\\
\mathbf{H}_{(p,q),(p^{\prime },q^{\prime })}^{0,d}+{\mathbf{H}}%
_{(p,q),(p^{\prime },q^{\prime })}^{0,c}+{\mathbf{H}}_{(p,q),(p^{\prime
},q^{\prime })}^{0,cv}+\mathbf{V(}i\mathbf{)}_{(p,q),(p^{\prime },q^{\prime
})}
\end{equation}
\begin{equation}
{\mathbf{H}}_{(p,q),(p^{\prime },q^{\prime })}^{0,d}=\delta _{p,p^{\prime
}}\delta _{q,q^{\prime }}%
\begin{pmatrix}
{\gamma }_{a}{(}\frac{2}{\Delta ^{2}}{+k}_{p}^{2}{+k}_{q}^{2}{)+E}_{g} & {0}
& {0} & {0} \\
{0} & {L}\frac{2}{\Delta ^{2}}{+M(k}_{p}^{2}{+k}_{q}^{2}{)} & {0} & {0} \\
{0} & {0} & {Lk}_{p}^{2}{+M(}\frac{2}{\Delta ^{2}}{+k}_{q}^{2}{)} & 0 \\
{0} & {0} & 0 & {Lk}_{q}^{2}{+M(}\frac{2}{\Delta ^{2}}{+k}_{p}^{2}{)}%
\end{pmatrix}%
\end{equation}
\begin{equation}
{\mathbf{H}}_{(p,q),(p^{\prime },q^{\prime })}^{0,c}=%
\begin{pmatrix}
{0} & {0} & {P}{\large \delta }_{q,q^{\prime }}\frac{4k_{p^{\prime }}p}{\pi
(p^{2}-p^{\prime 2})}{\large \delta }_{p+p^{\prime },odd} & {P}{\large %
\delta }_{p,p^{\prime }}\frac{4k_{q^{\prime }}q}{\pi (q^{2}-q^{\prime 2})}%
{\large \delta }_{q+q^{\prime },odd} \\
{0} & {0} & {0} & {0} \\
{\large -}{P}{\large \delta }_{q,q^{\prime }}\frac{4k_{p^{\prime }}p}{\pi
(p^{2}-p^{\prime 2})}{\large \delta }_{p+p^{\prime },odd} & {0} & {0} & 0 \\
{\large -}{P}{\large \delta }_{p,p^{\prime }}\frac{4k_{q^{\prime }}q}{\pi
(q^{2}-q^{\prime 2})}{\large \delta }_{q+q^{\prime },odd} & {0} & 0 & 0%
\end{pmatrix}%
\end{equation}
\begin{equation}
{\mathbf{H}}_{(p,q),(p^{\prime },q^{\prime })}^{0,cv}=-{N}\left( \frac{%
4k_{p^{\prime }}p}{\pi (p^{2}-p^{\prime 2})}\right) \left( \frac{%
4k_{q^{\prime }}q}{\pi (q^{2}-q^{\prime 2})}\right) {\delta }_{p+p^{\prime
},odd}{\delta }_{q+q^{\prime },odd}%
\begin{pmatrix}
{0} & {0} & {0} & {0} \\
{0} & {0} & {0} & {0} \\
{0} & {0} & {0} & 1 \\
{0} & {0} & 1 & {0}%
\end{pmatrix}%
\end{equation}
\normalsize
where ${\large \delta }$ is the Kronecker delta and ${\large \delta }%
_{p+p^{\prime },odd}=\{1,$ if $p+p^{\prime }=odd$, otherwise 0$\}$ and
\begin{equation}
\mathbf{V}(i)_{\{p,q\},\{p^{\prime },q^{\prime }\}}=\mathbf{I}_{4x4}\frac{4}{L_{y}L_{z}}%
\int\limits_{0}^{L_{y}}\int\limits_{0}^{L_{z}}\sin (k_{p^{\prime }}y)\sin
(k_{q^{\prime }}z)\phi (x=x_{i};y,z)\sin (k_{p}y)\sin (k_{q}z)dydz
\end{equation}
is orthogonal transformation to $\left\vert j,p,q,i\right\rangle $ of
Hartree $\phi (x=x_{i};y,z)$ \ potential (which is obtained self-consistently solving NEGF-Poisson equation), with $\mathbf{I}_{4x4}$ being the 4x4 identity matrix. Similarly, the inter-layer coupling matrix can be written in similar manner
\begin{equation}
\mathbf{W}_{(p,q),(p^{\prime },q^{\prime })}=\mathbf{W}_{(p,q),(p^{\prime
},q^{\prime })}^{d}+\mathbf{W}_{(p,q),(p^{\prime },q^{\prime })}^{c}\\
\end{equation}
, where
\footnotesize
\begin{equation}
\mathbf{W}_{(p,q),(p^{\prime },q^{\prime })}^{d}=\delta _{p,p^{\prime
}}\delta _{q,q^{\prime }}
\begin{pmatrix}
\frac{-{\gamma }_{a}}{\Delta ^{2}} & {\frac{{P}}{2\Delta }} & {0} & {0} \\
{\frac{-{P}}{2\Delta }} & \frac{-{L}}{\Delta ^{2}} & {0} & {0} \\
{0} & {0} & \frac{-{M}}{\Delta ^{2}} & 0 \\
{0} & {0} & 0 & \frac{-{M}}{\Delta ^{2}}%
\end{pmatrix}
\end{equation}
\begin{equation}
\mathbf{W}_{(p,q),(p^{\prime },q^{\prime })}^{c}=%
\begin{pmatrix}
{0} & 0 & {0} & {0} \\
0 & {0} & \frac{-{N}}{2\Delta }\frac{4k_{p^{\prime }}p}{\pi (p^{2}-p^{\prime
2})}{\large \delta }_{p+p^{\prime },odd}{\delta }_{q,q^{\prime }} & \frac{-{N%
}}{2\Delta }\frac{4k_{q^{\prime }}q}{\pi (q^{2}-q^{\prime 2})}{\large \delta
}_{q+q^{\prime },odd}{\delta }_{p,p^{\prime }} \\
{0} & \frac{-{N}}{2\Delta }\frac{4k_{p^{\prime }}p}{\pi (p^{2}-p^{\prime 2})}%
{\large \delta }_{p+p^{\prime },odd}{\delta }_{q,q^{\prime }} & {0} & 0 \\
{0} & \frac{-{N}}{2\Delta }\frac{4k_{q^{\prime }}q}{\pi (q^{2}-q^{\prime 2})}%
{\large \delta }_{q+q^{\prime },odd}{\delta }_{p,p^{\prime }} & 0 & 0%
\end{pmatrix}%
\end{equation}
\normalsize
\bigskip

One should mention that $k_{p},k_{q}$ form rectangular grid. Further simplification such as Hamiltonian size reduction in  \textbf{k} $\cdot${} \textbf{p} basis by taking only $k_{p},k_{q}$ vectors inside the circle\cite{shinMain}can be done to minimize memory usage and
computational power . Moreover, one can get further matrix size
reduction by employing the mode-space approach. Mode-space is crucial for the
recursive algorithm in NEGF implementation and charge distribution
construction. It was shown\cite{NonlSelf} that in case of electron-photon interaction
one cannot easily use recursive approach since self-energies are highly
non-local and in this case one has to take more off-diagonal blocks. In other words, more correlations between electron Green's functions  have to be kept when dealing with electron photon interaction.

\subsection{Electron-photon interaction. Monochromatic excitation.}
The electron-photon interaction part of Hamiltonian reads as
\begin{equation}
\label{eq:AP}
H_{e-ph}=-\frac{e}{m_{0}}\mathbf{A\cdot p}
\end{equation}
where the photon field is quantized and is given by
\begin{equation}
\mathbf{A}=\sum\limits_{\lambda ,\mathbf{q}}\left[ \mathbf{A}_{0}(\lambda ,%
\mathbf{q)b}_{\lambda ,\mathbf{q}}e^{i\omega _{\lambda }t}+\mathbf{A}%
_{0}(\lambda ,-\mathbf{q)b}_{\lambda ,-\mathbf{q}}^{\dagger }e^{-i\omega
_{\lambda }t}\right] e^{i\mathbf{qr}}
\end{equation}
\begin{equation}
\mathbf{A}_{0}(\lambda ,\mathbf{q)=e}_{\lambda ,\mathbf{q}}\sqrt{\frac{\hbar^{2}}{2\epsilon_{0} E_{\lambda }V}}
\end{equation}
where $\mathbf{b}_{\lambda ,\mathbf{q}}^{\dagger },\mathbf{b}_{\lambda ,%
\mathbf{q}}$ are the photon creation and annihilation operators,
respectively, $\mathbf{e}_{\lambda ,\mathbf{q}}$- is the polarization vector, $%
\mathbf{q\ }$- is the photon wavevector and $\lambda $- is the photon energy. Sum is over
all photon wavectors and energies.
where $V$ is the absorbing volume. The incident photon flux is related to photon occupation
number via
\begin{equation}
\Phi _{_{\lambda }}=\frac{N_{_{\lambda }}c}{V\sqrt{\mu \varepsilon }}=\frac{%
I_{_{\lambda }}}{E_{\lambda }}
\end{equation}
where $I_{_{\lambda }}$ - is the intensity of the EM field and $c$ - is the speed of
light. Equation (\ref{eq:AP}) in the second quantized form can be written as
\begin{equation}
\label{eq:HamEph}
H_{e-ph}=-\sum\limits_{(j,p,q,i),(j^{\prime },p^{\prime },q^{\prime
},i^{\prime })}\left\langle j^{\prime },p^{\prime },q^{\prime
},i^{\prime }\right\vert \mathbf{A\cdot p}\left\vert j,p,q,i\right\rangle
c_{j^{\prime },p^{\prime },q^{\prime },i^{\prime }}^{\dagger
}c_{j,p,q,i}(be^{i\omega _{\lambda }t}+b^{\dagger }e^{-i\omega _{\lambda }t})
\end{equation}
with $c_{j^{\prime },p^{\prime },q^{\prime },i^{\prime }}^{\dagger }$ -
being electron creation operator in the state symmetry $j^{\prime }$,
transverse subband $\{p^{\prime },q^{\prime }\}$, and position $x_{i^{\prime}}$ and
$c_{j,p,q,i}$ - being electron destruction operator in the state  of symmetry $j$,
transverse subband ${\{p,q\}}$, and position $x_{i}.$

Carrying out explicitly matrix element of (\ref{eq:HamEph}) in dipole approximation with wire dimensions much smaller than a wavelength ($\mathbf{qr}<<1)$ and taking into account only inter-subband excitations(CB-VB), i.e. considering only CB-VB transitions, we arrive at
\begin{equation}
M_{\{j,p,q,i\},\{j^{\prime },p^{\prime },q^{\prime },i^{\prime }\}}=
\left\langle j^{\prime },p^{\prime },q^{\prime },i^{\prime }\right\vert
\mathbf{A\cdot p}\left\vert j,p,q,i\right\rangle =\delta _{p,p^{\prime
}}\delta _{q,q^{\prime }}\delta _{i,i^{\prime }}\mathbf{A\cdot p}%
_{i,\{j,j^{\prime }\}}
\end{equation}
where
\begin{subequations}
\begin{align}
\mathbf{p}_{i,\{j,j^{\prime }\}}=\{\left\langle S\uparrow \downarrow
\right\vert p_{x}\left\vert X\uparrow \downarrow \right\rangle ,\left\langle
S\uparrow \downarrow \right\vert p_{y}\left\vert Y\uparrow \downarrow
\right\rangle ,\left\langle S\uparrow \downarrow \right\vert p_{z}\left\vert
Z\uparrow \downarrow \right\rangle \}\\
\left\langle S\uparrow \downarrow \right\vert p_{x}\left\vert X\uparrow
\downarrow \right\rangle =\left\langle S\uparrow \downarrow \right\vert
p_{y}\left\vert Y\uparrow \downarrow \right\rangle =\left\langle S\uparrow
\downarrow \right\vert p_{z}\left\vert Z\uparrow \downarrow \right\rangle =\frac{m_{0}}{\hbar}iP
\end{align}
\end{subequations}

Total Matrix ($\mathbf{M}_{e-ph}$) becomes

\begin{equation}
\mathbf{M}_{e-ph}^{\{x,y,z\}}=A_{0}\frac{m_{0}}{\hbar }
\begin{pmatrix}
\mathbf{M}_{(1,1)(1,1)}^{\{x,y,z\}} & 0 & ... & ... & 0 \\
0 & 0 & ... & ... & 0 \\
... & ... & 0 & ... & ... \\
... & 0 & \mathbf{M}_{(p,q)(p,q)}^{\{x,y,z\}} & ... & ...\\
... & 0 & 0 & ... & 0 \\
0 & ... & ... & 0 & \mathbf{M}^{\{x,y,z\}}_{(N_{p}N_{q})(N_{p}N_{q})}
\end{pmatrix}
\end{equation}
where,
\begin{equation}
\mathbf{M}_{(p,q)(p,q)}^{x}=
\begin{pmatrix}
0 & iP & 0 & 0 \\
-iP & 0 & 0 & 0 \\
0 & 0 & 0 & 0 \\
0 & 0 & 0 & 0%
\end{pmatrix}%
\end{equation}
\begin{equation}
\mathbf{M}_{(p,q)(p,q)}^{y}=%
\begin{pmatrix}
0 & 0 & iP & 0 \\
0 & 0 & 0 & 0 \\
-iP & 0 & 0 & 0 \\
0 & 0 & 0 & 0%
\end{pmatrix}%
\end{equation}
\begin{equation}
\mathbf{M}_{(p,q)(p,q)}^{z}=%
\begin{pmatrix}
0 & 0 & 0 & iP \\
0 & 0 & 0 & 0 \\
0 & 0 & 0 & 0 \\
-iP & 0 & 0 & 0%
\end{pmatrix}%
\end{equation}
\begin{equation}
\mathbf{M}_{(p,q)(p,q)}^{\{l.c.\}}=%
\begin{pmatrix}
0 & iP & iP & iP \\
-iP & 0 & 0 & 0 \\
-iP & 0 & 0 & 0 \\
-iP & 0 & 0 & 0%
\end{pmatrix}%
\end{equation}

where {x,y,z} stands for either x,y or z EM field polarization, l.c. stands for the linear polarization which is linear combination of the x,y and z axis.

\subsection{NEGF and Self-Energies}
Green's functions are assumed to be in steady state with electron Green's function
being at zero temperature (although temperature comes via Fermi levels) and photon Green's functions being unperturbed
by electronic elementary excitations. Within Keldysh formalism the  Dyson's equations of motion for the
electronic Green's functions in matrix notation are given by
\begin{subequations}
\begin{align}
\mathbf{G}^{R}(E)=\left( (E+i\gamma)\mathbf{I} -{\mathbf{H}}_{tot}(E)-%
\mathbf{\Sigma }^{B}\mathbf{(}E)-\mathbf{\Sigma }_{e-ph}%
\mathbf{(}E)\right) ^{-1}\\
\mathbf{G}^{<}(E)=\mathbf{G}^{R}(E)\left\{
\mathbf{\Sigma }^{<,B}\mathbf{(}E)+\mathbf{\Sigma }_{e-ph}^{<}%
\mathbf{(}E)\right\} \mathbf{G}^{A}(E)\\
\mathbf{G}^{A}(E)=[\mathbf{G}^{R}(E)]^{\dag }\\
\mathbf{G}^{>}(E)=\mathbf{G}^{R}(E)-\mathbf{G}%
^{A}(E)+\mathbf{G}^{<}(E)
\end{align}
\end{subequations}
where $\mathbf{\Sigma }^{B}\mathbf{(}E)$ is the boundary
self-energy, which incorporates effect of semi-infinite contact(coupling to
contacts). Contacts are are assumed to be with equilibrium with right and left leads respectively  and are  perfect absorbers\cite{contacts}.   $\mathbf{\Sigma}_{e-ph}(E)$ is the electron-photon self-energy describing electron-photon interaction, where
\begin{equation}
\mathbf{\Sigma }^{B}(E)\equiv\mathbf{\Sigma }^{B}\mathbf{(}E)=\\
\begin{pmatrix}
\mathbf{\Sigma }_{L}^{B}(E) & 0 & ... & 0 \\
0 & 0 & ... & 0 \\
... & ... & ... & 0 \\
0 & ... & 0 & \mathbf{\Sigma }_{R}^{B}(E)%
\end{pmatrix}%
\end{equation}
where $\mathbf{\Sigma }_{L,R}^{B}(E)$ are the block matrices  of
size $4N_{p}N_{q}$ that are related to surface Green's functions via
\begin{subequations}
\begin{align}
\mathbf{\Sigma}_{L}^{B}(E)=%
\mathbf{W}\mathbf{g}_{L}(E)\mathbf{W}^{\dag }\\
\mathbf{\Sigma}_{R}^{B}(E)=%
\mathbf{W}^{\dag }\mathbf{g}_{R}(E)\mathbf{W}
\end{align}
\end{subequations}
where,
\begin{subequations}
\label{eq:GreenSf}
\begin{align}
\mathbf{g}_{L}(E)=[E-\mathbf{H}_{1}-\mathbf{\mathbf{W}^{\dag }g}_{L}(E)\mathbf{W}]^{-1}\\
\mathbf{g}_{R}(E)=[E-\mathbf{H}_{N_{x}}-\mathbf{W}\mathbf{g}_{R}(E)\mathbf{W}^{\dag }]^{-1}
\end{align}
\end{subequations}
are surface Green's functions corresponding to left and right lead, respectively. Equations on the $\mathbf{g}_{L}(E),\mathbf{g}_{R}(E)$ are matrix
quadratic equations. There are many ways of calculating the solution to (\ref{eq:GreenSf}). Simplest solution is just straightforward iteration, although this is very slowly converging process. Therefore, we have adopted the improved version of Anderson mixing \cite{Anderson} which is also simple in implementation.
Lesser(in-scattering) boundary self energy in case of equilibrated contacts is given by
\begin{subequations}
\begin{align}
\mathbf{\Sigma }^{<,B}_{L,R}(E)=i\mathbf{\Gamma}_{L,R}(E)f_{L,R} \\
\mathbf{\Gamma}_{L,R}(E)=i(\mathbf{\Sigma^{B}}_{L,R}(E)-\mathbf{\Sigma^{B,\dag}}_{L,R}(E))
\end{align}
\end{subequations}
where $f_{L,R}$ are the Fermi levels at the left and right lead respectively, and $\mathbf{\Gamma }_{L,R}$ is  the level broadening.

Light-matter interaction leads to electron-hole pair generation and electron-hole recombination by absorbing/emitting a photon. This process is inelastic,  and in general is phase-breaking. In order to incorporate this interaction into NEGF formalism in the first order Born-approximation(one-photon processes) one has to utilize Wick's theorem and Langreth contour rules as it was done in several works\cite{LakeEn,SteigerPhD} and in the original Henrickson's\cite{henrickson} papers.

Most self-energies of this form, including electron-photon,(fermion-boson interaction in the limit one elementary exciation) are current conserving\cite{MahanRev}.  In order to achieve current conservation one has to utilize self consistency among Green's functions and self-energies - in other words use self-consistent Born approximation(SCBA) or one can use current conserving schemes using Non-self consistent Born Approximation  described in Lake\cite{Lake} et.al. A detailed derivation of the self-consistent Born approximation approach is given in the work of Jiang et al.\cite{paper3}. Lesser and greater parts, $\mathbf{\Sigma}_{E-ph}^{<,>}$ are given by
\begin{subequations}
\begin{align}
\mathbf{\Sigma }_{e-ph}^{<,>}(x,x^{\prime },E)=\mathbf{\Sigma }%
_{e-ph}^{<,>,abs}(E)+\mathbf{\Sigma }_{e-ph}^{<,>,em}(E)+\mathbf{\Sigma }_{e-ph}^{<,>,sp}(E)\\
\mathbf{\Sigma }_{e-ph}^{<,>,abs}(E)=N_{\lambda }%
\mathbf{M}_{e-ph}\mathbf{G}^{<,>}(E\mp\hbar \omega )\mathbf{M}%
_{e-ph}\\
\mathbf{\Sigma }_{e-ph}^{<,>,em}(E)=N_{\lambda }\mathbf{%
M}_{e-ph}\mathbf{G}^{<,>}(E\pm\hbar \omega )\mathbf{M}_{e-ph}\\
\mathbf{\Sigma }_{e-ph}^{<,>,sp}(E)=\mathbf{M}%
_{e-ph}(\int\limits_{E_{\min }}^{E_{\max }}d(\hbar \omega _{\gamma })%
\mathbf{G}^{<,>}(E)\pm\hbar \omega _{\gamma }))\mathbf{M}_{e-ph}
\end{align}
\end{subequations}
where $\mathbf{\Sigma }_{e-ph}^{<,>,abs},\mathbf{\Sigma }_{e-ph}^{<,>,em},\mathbf{\Sigma }_{e-ph}^{<,>,sp}$ are the self energies associated with photon absorption, stimulated emission and spontaneous emission, respectively. The derivation is very similar to the work of Jiang et al.\cite{paper2} One should note that spontaneous emission term is integrated over broad energy range in CB and VB energy regions and is only dependent on joint density of states and occupation numbers at energies which differs by photon energy. $ E_{min}, E_{max}$ are the minimal and maximal photon energies dictated by material and device parameters. $\mathbf{M}_{e-ph}$ is the full electron-photon interaction Hamiltonian  in the basis $\left\vert j,p,q,i\right\rangle $.
Strictly speaking, one has to be careful considering $\mathbf{M}_{e-ph}$ since originally it couples only bulk CB-VB bands. In other words, if one wants to consider inter-subband excitations such as CB-CB or VB-VB (either within CB or VB manifolds),  the $\mathbf{M}_{e-ph}$  has to be modified accordingly to include intraband coupling in the original bulk model since one 3D band gives raise to many 1D subbands. In case of short-channel devices under certain biases the spontaneous term is assumed to be small\cite{Aerb2} and as will be shown later can be neglected. Real part of the retarded $\mathbf{\Sigma }_{e-ph}$ self energy is neglected since it leads just to energy renormalization\cite{Aerb2}, and only imaginary part of the $\mathbf{\Sigma }_{e-ph}$ is important and given by ( $(x,x^{\prime})$ notation is omitted throughout for simplicity)
\begin{equation}
Im(\mathbf{\Sigma}_{e-ph}(E))=\frac{1}{2}(\mathbf{\Sigma }%
_{e-ph}^{>}(E)-\mathbf{\Sigma }_{e-ph}^{<}(E))
\end{equation}
\subsection{Mode space and Physical quantities}
In case of mode-space\cite{shinMain} transformation one defines mode $m$ in the following manner
\begin{equation}
\Phi _{m}(x_{i},y_{j},z_{k})=\sum\limits_{j,p,q}\alpha _{j,p,q}^{m}(i)\left\vert p,q,j\right\rangle
\end{equation}
which satisfies 2D-sliced Schrodinger equation at slice $i$
\begin{equation}
(\mathbf{H}_{i}+\mathbf{W}+\mathbf{W}^{\dag}) \Phi _{m}(x_{i},y_{j},z_{k})=E_{m}\Phi _{m}(x_{i},y_{j},z_{k})
\end{equation}
Original eigenfunction of (\ref{eq:PSI}) is given in terms of modes as
\begin{equation}
\left\vert \Psi \right\rangle=\sum\limits_{m}Y_{m}(x_{i})\Phi _{m}(x_{i},y_{j},z_{k})
\end{equation}

In order to have self-consistent  NEGF with Poisson one has to compute 3D electron density in the real space representation. An incomplete\cite{luisierMode} calculation consists of writing 3D electron density in real space \cite{modeSpace} neglecting the mode correlation effects as
\begin{eqnarray}
n^{rs}_{3D}(i,j,k) &=& \frac{-2i}{\Delta\Delta_{y}\Delta_{z}}%
\int \frac{dE}{2\pi }(\mathbf{U}_{K}\mathbf{U}_{M}\mathbf{G}%
^{<,ms}(x,x^{\prime },E)\mathbf{U}_{M}^{\dag }\mathbf{U}_{K}^{\dag
})_{\{i,j,k\},\{i,j,k\}}\simeq \nonumber \\
&\simeq&\alpha\sum\limits_{n=1}^{N_{m}}%
\int \frac{dE}{2\pi }G_{(i,n),(i,n)}^{<,ms}(E)\left%
\vert \sum\limits_{j,p,q}\sin (k_{p}y_{j})\sin (k_{q}z_{k})\alpha_{j,p,q}^{n}(i)\right\vert ^{2}
\end{eqnarray}

where $\alpha=\frac{-2i}{\Delta\Delta_{y}\Delta _{z}} \frac{4}{N_{y}N_{z}}$, $rs,ms$ superscripts stand for the real-space and mode-space representations respectively. $G_{(i,n),(i,n)}^{<,ms}$ stands for the diagonal matrix element of mode $n$ at  block $i$. $\mathbf{U}_{K}, \mathbf{U}_{M} $ are unitary transformation matrices\cite{shinMain} defined as block diagonal matrices
built from $\mathbf{U}_{k}, \mathbf{U}_{m}(i) $ respectively, where

\footnotesize
\begin{equation}
\mathbf{U_{k}}=\tfrac{2}{\sqrt{N_{y}N_{z}}}%
\begin{pmatrix}
\sin {(k_{1}y_{1})}\sin {(k_{1}{z}_{1})}\mathbf{I}_{4x4} & ... &
 \sin {(k_{1}z_{1})}\sin {(k_{N_{q}}z_{1})}\mathbf{I}_{4x4} & ... & \sin {(k_{N_{p}} y_{1})} \sin {k_{N_{q}}z_{1}}
 \mathbf{I}_{4x4} \\
\sin {(k_{1}y_{1})} \sin {(k_{1}z_{2})} \mathbf{I}_{4x4} & ... &
\sin {(k_{1}y_{1})} \sin {(k_{N_{q}}z_{2})} \mathbf{I}_{4x4} & ... &
\sin {(k_{N_{p}}y_{1})} \sin {(k_{N_{q}}z_{2})} \mathbf{I}_{4x4} \\
... & \ddots  & ... & ... & ... \\
\sin {(k_{1}y_{1})} \sin {(k_{1}z_{N})}\mathbf{I}_{4x4} & ... &
\sin {(k_{1}y_{1})} \sin {(k_{N_{q}}z_{N})}\mathbf{I}_{4x4} & ... &
\sin {(k_{N_{p}}y_{1})} \sin {k_{N_{q}}z_{N})} \mathbf{I}_{4x4} \\
... & ... & ... & \ddots  & ... \\
\sin {(k_{1}y_{M})} \sin {(k_{1}z_{N})}\mathbf{I}_{4x4} & ... & ... & ... & \sin {(k_{N_{p}}y_{M})}
\sin {(k_{N_{q}}z_{N})}\mathbf{I}_{4x4}%
\end{pmatrix}%
\end{equation}
\normalsize
is the size of $(4NM)\times(4N_{p}N_{q})$ and

\begin{equation}
\mathbf{U_{m}}(i)=%
\begin{pmatrix}
\alpha _{1,1,1}^{1}(i) & \alpha _{1,1,1}^{2}(i) & ... & ... & \alpha
_{1,1,1}^{N_{m}}(i) \\
\alpha _{2,1,1}^{1}(i) & \alpha _{2,1,1}^{2}(i) & ... & ... & ... \\
... & ... & ... & ... & ... \\
... & ... & \alpha _{j,N_{p},N_{q}}^{n}(i) & ... & ... \\
\alpha _{4,N_{p},N_{q}}^{1}(i) & \alpha _{4,N_{p},N_{q}}^{2}(i) & ... & ...
& \alpha _{4,N_{p},N_{q}}^{N_{m}}(i)%
\end{pmatrix}
\end{equation}
is the size of $(4N_{p}N_{q})\times N_{m}$

Current flowing between layers $i$ ,and $i+1$ can be written as
\begin{equation}
I_{x_{i}\rightarrow x_{i+1}}=\frac{2e}{\hbar }\int \frac{dE}{2\pi }tr\{%
\mathbf{W}^{ms}\mathbf{G}_{i,i+1}^{<,ms}-\mathbf{W}^{\dag ,ms}\mathbf{G}%
_{i+1,i}^{<,ms}\}
\end{equation}

Similar approach has been applied in the study of thermal expansion of single-wall carbon nanotubes and grapheme sheets\cite{paper1}

\section{Numerical Details}
\begin{figure}[h]
\centering
\includegraphics[width=300.0px]{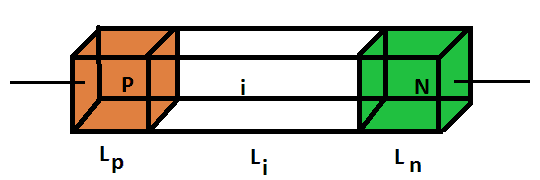}
\caption{nanosized p-i-n diode}
\label{fig:pin}
\end{figure}

The device under study is a p-i-n structure and is depicted in Fig. \ref{fig:pin}. The device is 42 nm long, with a square cross-section of 10nm x 10nm. The doping on both the $n$ and $p$ ends is assumed to be  $3.2*10^{18}$  $cm^{-3}$. Furthermore, length of the $p$ and $n$ region was set  $L_{p}=L_{n}=12$nm and inter-layer spacing $\Delta=0.3$nm .   Current conserving grid was chosen\cite{SteigerPhD} as $\Delta E=E_{\lambda}/N_{ph}$ with total number of energy grid points $N_{tot}=Int((|E_{1}|+|E_{2}|)/\Delta E)$ with $E_{1}, E_{2} $ being conduction and valence band cut-off energies chosen accordingly to the region of interest. $N_{ph}$ defines by how many energy points separated $E$ and $E+\hbar \omega$. $N_{tot}$ varied between 800 and 2000 points to make sure convergence is achieved in energy space. $N_{x}$ was set to 140 points. The potential profile is assumed to be uniform in the cross-sectional area.  1D potential profiles and Fermi-levels were obtained by nextnano simulator \cite{nextnano} with the parameters being $E_g$=1.42eV, $m_c$=0.067$m_e$, $m_h$=0.082$m_e$ where parameters are bandgap, effective conduction mass, effective valence mass(light hole) respectively. Although, strictly speaking there is no physical justification for this, but it does not affect the physical picture except consideration of the boundary effects in which we are not interested at the moment. 1D Potential profiles and Quasi-Fermi levels were fed into optical NEGF simulator based on 2 subband model, which is written as

\begin{equation}
\mathbf{H}_{2x2}(E)=%
\begin{pmatrix}
E_{g}+\frac{\hbar ^{2}}{2m_{0}}\frac{\gamma _{a}}{\Delta ^{2}} & 0 \\
0 & \frac{\hbar ^{2}}{2m_{0}}\frac{\gamma _{l}}{\Delta ^{2}}%
\end{pmatrix}
\end{equation}

\begin{equation}
\mathbf{W}_{2x2}(E)=%
\begin{pmatrix}
-\frac{\hbar ^{2}}{2m_{0}}\frac{\gamma _{a}}{\Delta ^{2}} & \frac{P}{2\Delta
} \\
-\frac{P}{2\Delta } & -\frac{\hbar ^{2}}{2m_{0}}\frac{\gamma _{l}}{\Delta
^{2}}
\end{pmatrix}%
\end{equation}

In order to avoid spurious solutions in $k$- space $(\mathbf{k}=(k_{x},-i\frac{\partial }{\partial y},-i\frac{\partial }{\partial z}))$ we took cross-sectional area such that condition on the envelope function is satisfied so, that plane-wave expansion lies in the first Brillouin-zone\cite{EnvelopeFirstBrill,Foreman95,Burt,CutoffK}
\begin{equation}
\frac{2\pi N_{p,q}^{max} }{L_{y,z}}\ll \frac{2\pi }{a}
\end{equation}
with $N_{p,q}^{max}\le L_{y,z}/a$. In addition  we set $\gamma_{c}=0$ with optimizing the parameters \cite{Foreman,Foreman75} $\gamma_{1},\gamma_{2},\gamma_{3}$ \cite{Parameters} such that bulk effective masses of hole and electrons are reproduced. Moreover, the original Hamiltonian can be modified to avoid spurious solutions \cite{Kolokolov}.  Going from  $k$- space representation to real-space representation $(-i\frac{\partial }{\partial x},-i\frac{\partial }{\partial y}, -i\frac{\partial }{\partial z} )$  with finite differences being the basis one has another source of spurious solutions\cite{real_spurious_meshing,FEM}. Such solutions can be avoided by using certain finite element basis\cite{FEM}. However in general,  there is no common remedy for this type of problem\cite{FEM}. In particular, to avoid this type of problem, one either chooses  inter-layer spacing $\Delta$ accordingly to the parameters $\gamma_{1},\gamma_{2},\gamma_{3}$ \cite{real_spurious_meshing}, or as  we did, fix the $\Delta$ and vary the parameters $\gamma _{a},\gamma _{l}$ to reproduce the bulk effective  masses of conduction band and light-hole bands (we have assumed that charge carrier effective masses of 10nm x 10nm are bulk values). Parameters after fitting are $\gamma _{a}=8,\gamma _{l}=-1$. SCBA computations are aborted once convergence is achieved by monitoring the norm of the total photocurrent $\int(I_{ph}^{e}+I_{ph}^{h})_{n+1}dE/\int(I_{ph}^{e}+I_{ph}^{h})_{n}dE<\epsilon $, where  $\epsilon$  was set to $10^{-4}$. In the computation of the spontaneous emission term, we have set $E_{min},E_{max}$ to be in the range of $E_{\lambda}\mp0.4E_{\lambda}$, for particular photon energy with $0.4E_{\lambda}$ term being chosen such that results are converged meanwhile minimizing the computational resources . Laser intensity is assumed to be 100 $W/cm^{2}$ unless specified otherwise. The whole structure is uniformly illuminated. EM field polarized along x-axis. Device structure is depicted on the Figure \ref{fig:pin}.

\section{Results and discussion}
\begin{figure}
\centering
\includegraphics[width=300.0px]{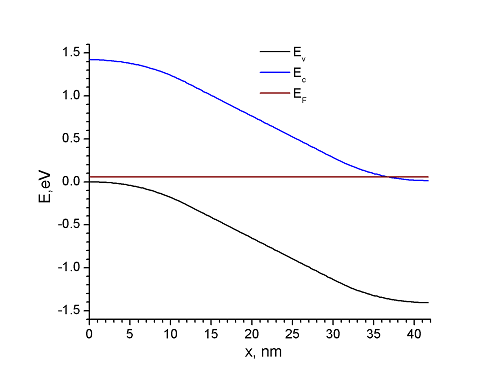}
\caption{Band profile for the conduction and valence band. Brown line indicates Fermi level.}
\label{fig:BandProfile}
\end{figure}

\begin{figure}
\centering
\includegraphics[width=300.0px]{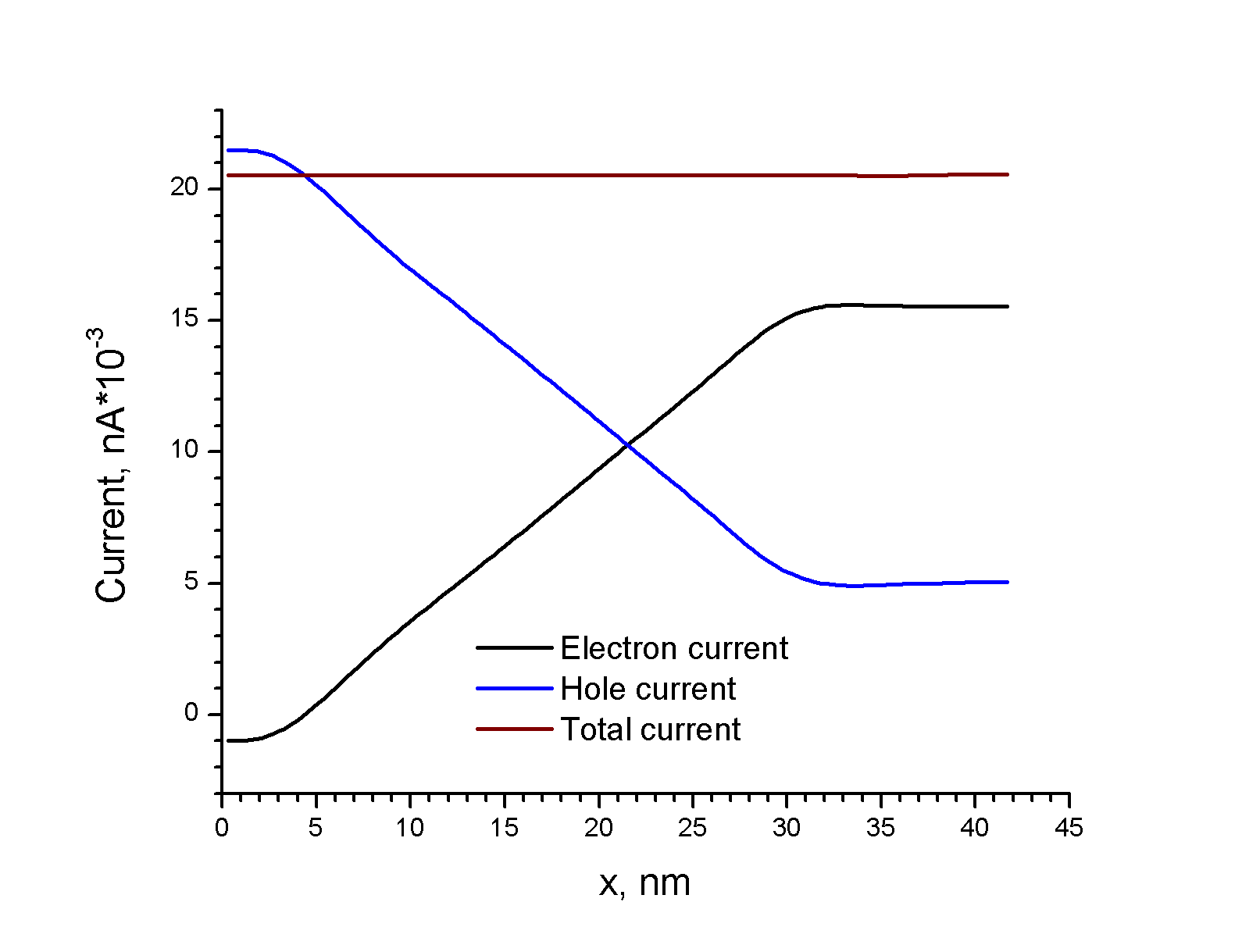}
\caption{Spatial distribution of hole and electron current at $E_{\lambda}=1.56 eV$ at zero bias
for p-i-n structure. One can see that total current is conserved }
\label{fig:SpatialCurr}
\end{figure}

\begin{figure}
\centering
\includegraphics[width=300.0px]{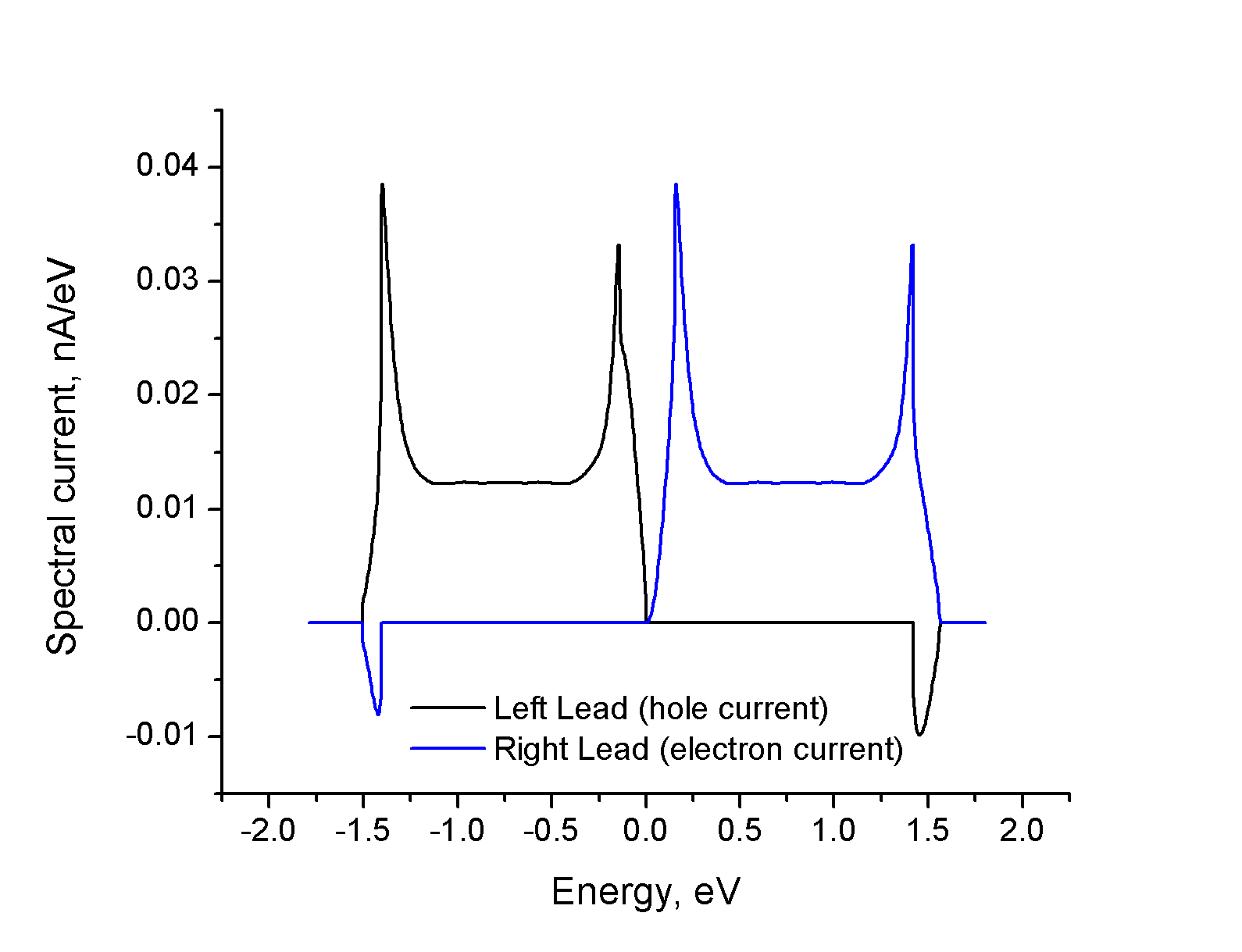}
\caption{Energetic current distribution over valence and conduction band at $E_{\lambda}=1.56 eV$ at zero bias for p-i-n structure.}
\label{fig:EnergDistrCurr}
\end{figure}

\begin{figure}
\centering
\includegraphics[width=300.0px]{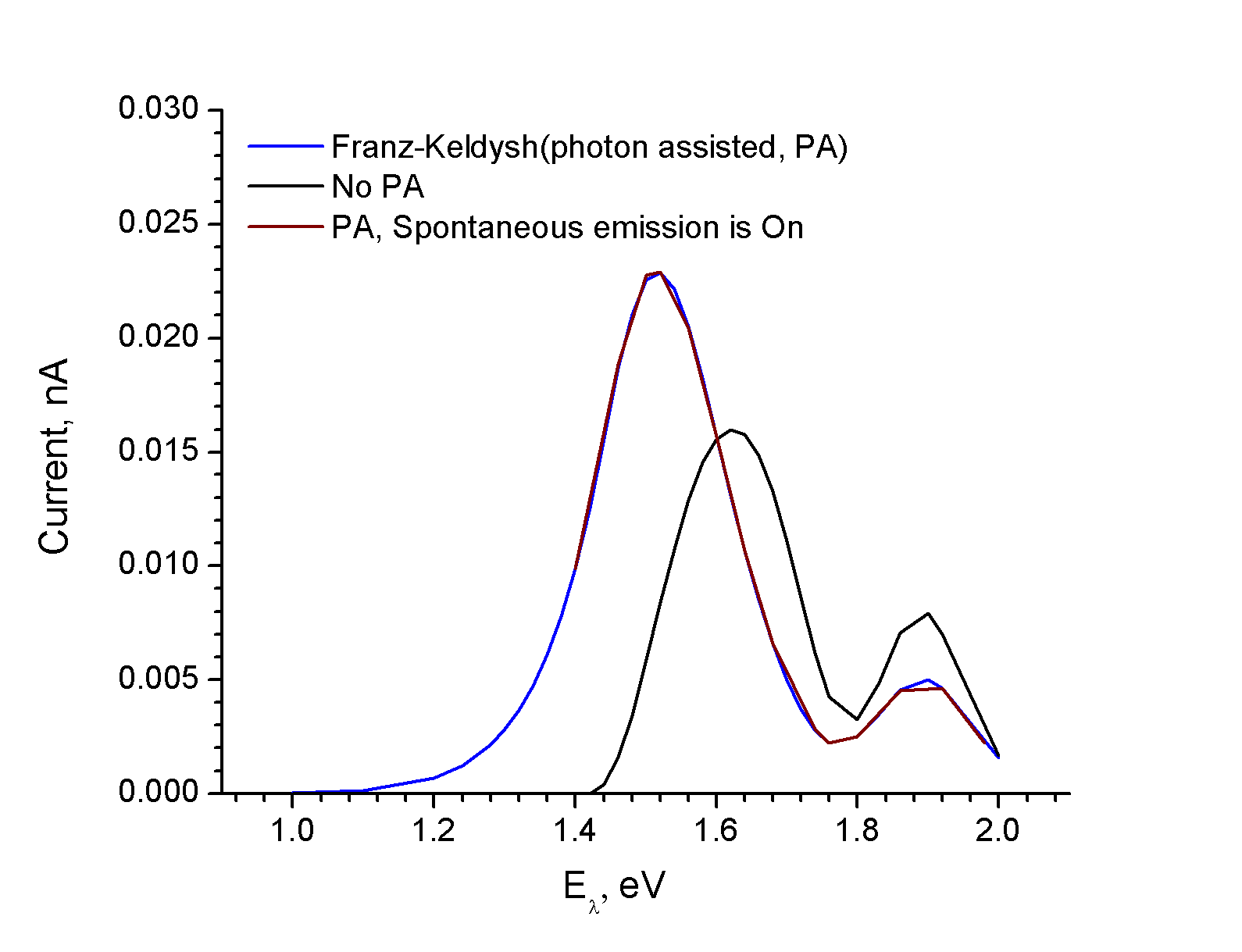}
\caption{Photoresponse at zero bias for p-i-n structure with and without photon-assisted (Franz-Keldysh effect) tunneling.}
\label{fig:PANOPA}
\end{figure}

\begin{figure}[!]
  \begin{center}
    \subfigure[LDOS]{\label{fig:LDOS-a}\includegraphics[scale=0.44]{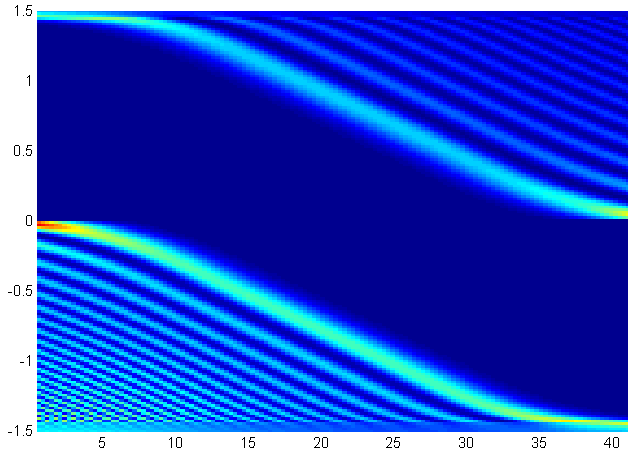}}
    \subfigure[LDOS cross sectioned]{\label{fig:LDOS-b}\includegraphics[scale=0.31]{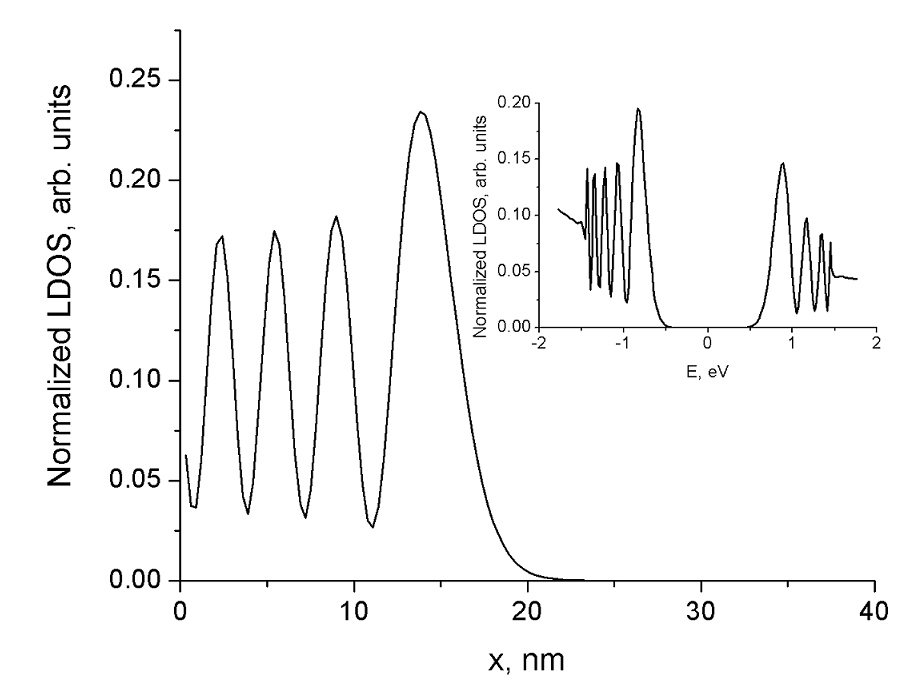}}
  \end{center}
  \caption{Local Density of States. a) LDOS plot (x-axis is position in nm, y-axis is energy in eV) b) Cross sections of the LDOS. Main graph is the spatial LDOS near the top of the Valence band. Inset is the cross section over energy coordinate near the middle of the device.}
  \label{fig:LDOS}
\end{figure}

Figure \ref{fig:BandProfile} depicts the conduction and valence band profiles. As can be seen, the built-in potential  is $V_{bi}=$1.4 V. In the calculation, we have used non-self-consistent Poisson profile,  which deviates from the self-consistent solution by less than 5\%\cite{Nanotube_Light} at a light intensity $I=10^{5}$  $W/cm^{2}$. Since the light intensity in our work is much smaller, therefore one expects even lower deviation from SCF Poisson profile.
Upon light illumination of the diode, the electron-hole pairs start to form, which are then separated by the electric field.
Figure \ref{fig:SpatialCurr} shows typical spatial hole and electron current distribution. One can see that total current is conserved, meanwhile hole current grows towards the $p$-contact and electron current as we move towards the $n$-contact.
To investigate it further we have calculated the current energetic distribution in the device, including left and right leads. Typical spectral current is shown in Fig.\ref{fig:EnergDistrCurr}. One clearly sees three regions contributing to the current, with the flat region corresponding to the channel current, and two peaks corresponding to the $p$-region and $n$-region currents(flowing just above and below conduction and valence band edges, respectively). The distribution is not symmetric with respect to $p$ and $n$ regions. Figure \ref{fig:PANOPA} shows photocurrent in the short circuit(sc) condition at different photon energies. One can see formation of two peaks in the Franz-Keldysh (photon-assisted tunneling or PA) regime and without it. The manifestation of the Franz-Keldysh effect is the non-zero current below fundamental bandgap. In PA regime the first peak is shifted towards lower energies due to non-zero DOS below the bandgap. It is seen that just interband approximation (zero DOS below CB and below VB edges respectively) significantly underestimates the photocurrent. The channel current grows as photon energy increases due to a greater availability of DOS. The local density of states (LDOS) has an oscillatory pattern both in spatial and energetic coordinates as seen in Fig.\ref{fig:LDOS} which forms due to incident $k_{+}$ and reflected $k_{-}$ electron waves. By keeping all off-band correlations\cite{AerbCoh1,HohCoh} in the Green's functions , i.e. elements such as $G_{x,x^{\prime },c,v}^{<}(E)$  and where spontaneous emission does not play significant role  with coherent light\cite{AerbCoh} source one has phase-coherent photo-response. This is automatically satisfied in our case, since we are working with the full rank of the matrix.


\begin{figure}
\centering
\includegraphics[width=300.0px]{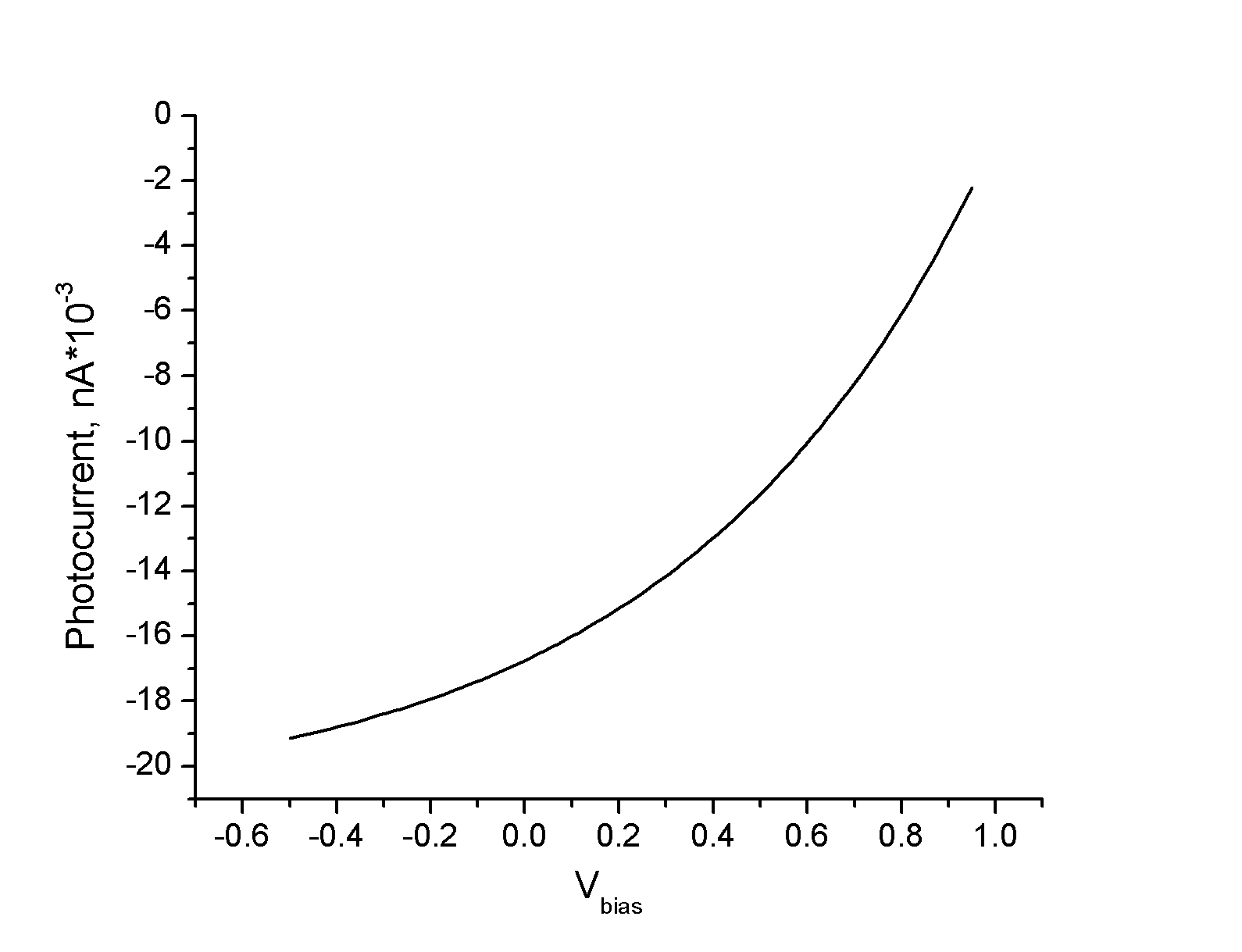}
\caption{IV characteristics at $E_{\lambda}=1.56 eV$. Please note that the current sign is take opposite of the original one (originally current flow taken positive from $p$ lead to $n$ lead). }
\label{fig:IV}
\end{figure}

\begin{figure}[!]
  \begin{center}
    \subfigure[$~E_\lambda=1.40$ eV]{\label{fig:2Ds-a}\includegraphics[scale=0.5]{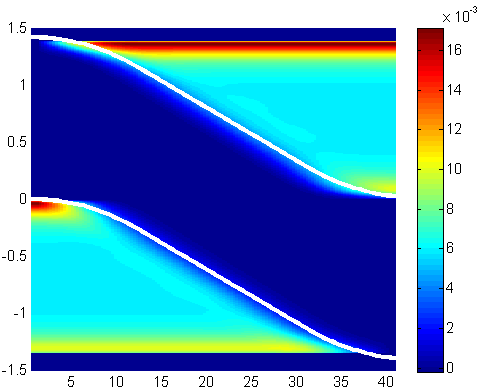}}
    \subfigure[$~E_\lambda=1.52$ eV]{\label{fig:2Ds-b}\includegraphics[scale=0.5]{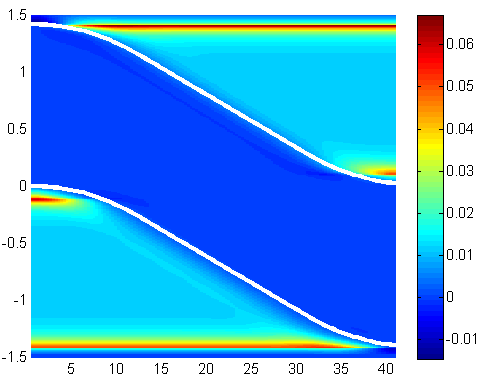}}
    \subfigure[$~E_\lambda=1.64$ eV]{\label{fig:2Ds-c}\includegraphics[scale=0.5]{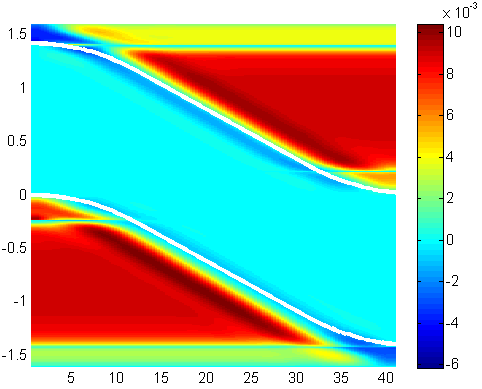}}
    \subfigure[$~E_\lambda=1.76$ eV]{\label{fig:2Ds-d}\includegraphics[scale=0.5]{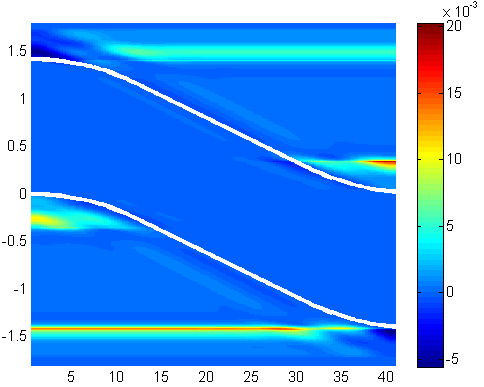}}
    \subfigure[$~E_\lambda=1.86$ eV]{\label{fig:2Ds-e}\includegraphics[scale=0.5]{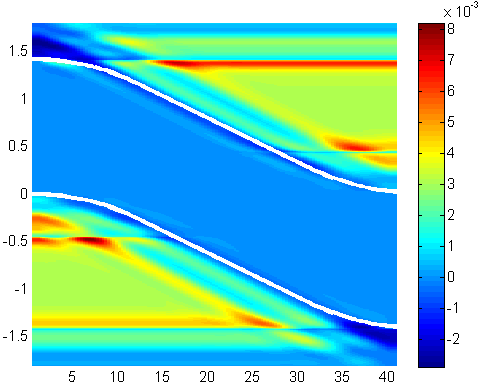}}
    \subfigure[$~E_\lambda=2.00$ eV]{\label{fig:2Ds-f}\includegraphics[scale=0.5]{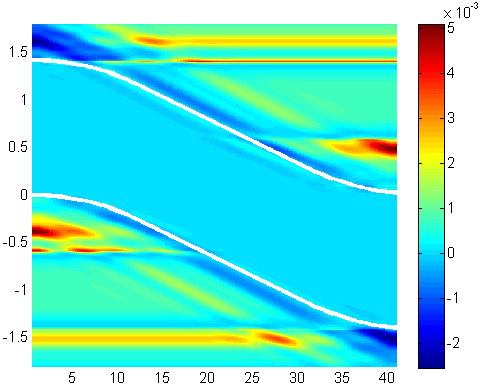}}
  \end{center}
  \caption{2D distribution spectral current for six different photon energies. x-axis units are nm, y-units are eV. Colormap units correspond to nA/eV }
  \label{fig:2Ds}
\end{figure}

\begin{figure}
  \begin{center}
    \subfigure[$~E_\lambda=1.64$ eV]{\label{fig:CBVB-a}\includegraphics[scale=0.3]{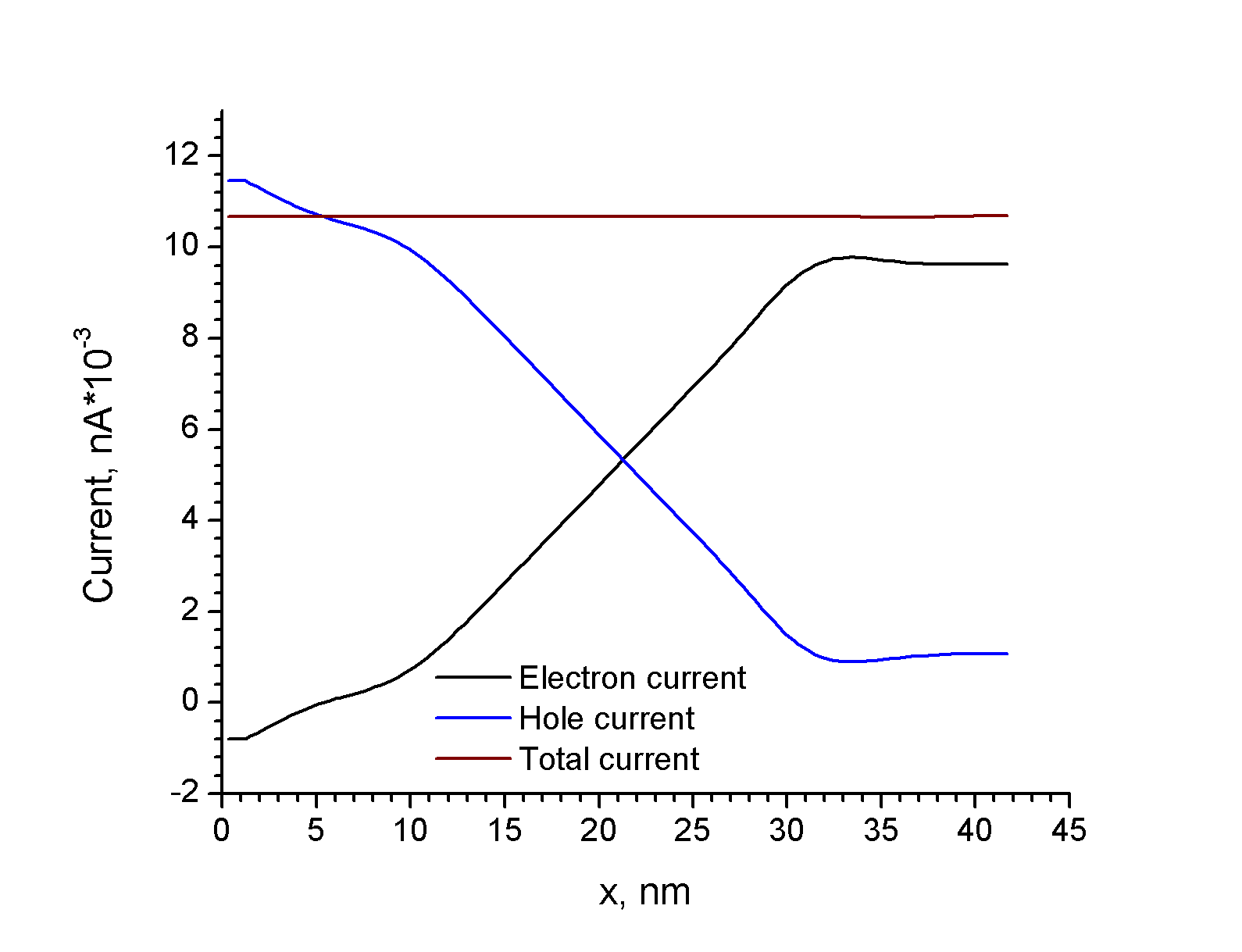}}
    \subfigure[$~E_\lambda=1.76$ eV]{\label{fig:CBVB-b}\includegraphics[scale=0.3]{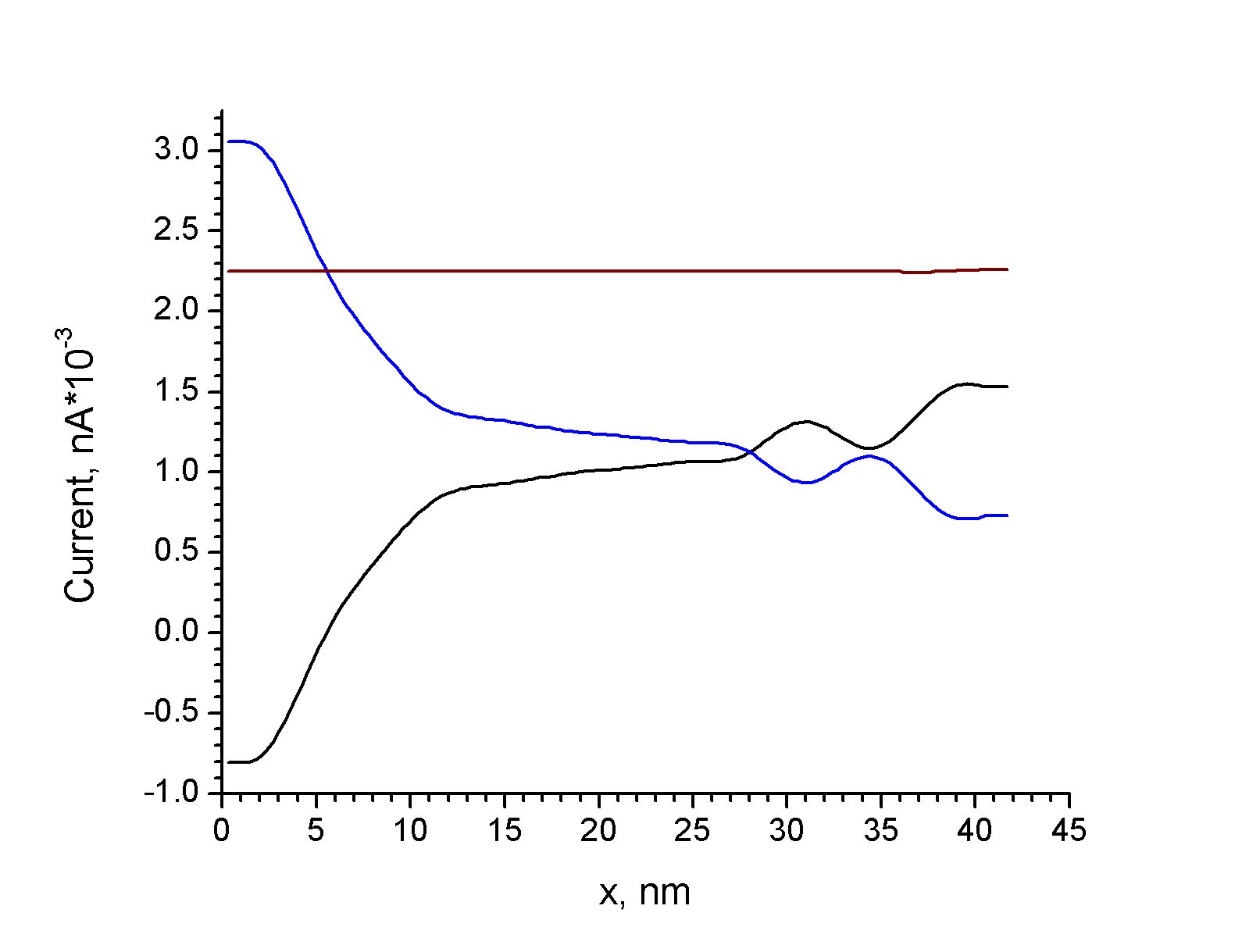}}
    \subfigure[$~E_\lambda=2.00$ eV]{\label{fig:CBVB-c}\includegraphics[scale=0.3]{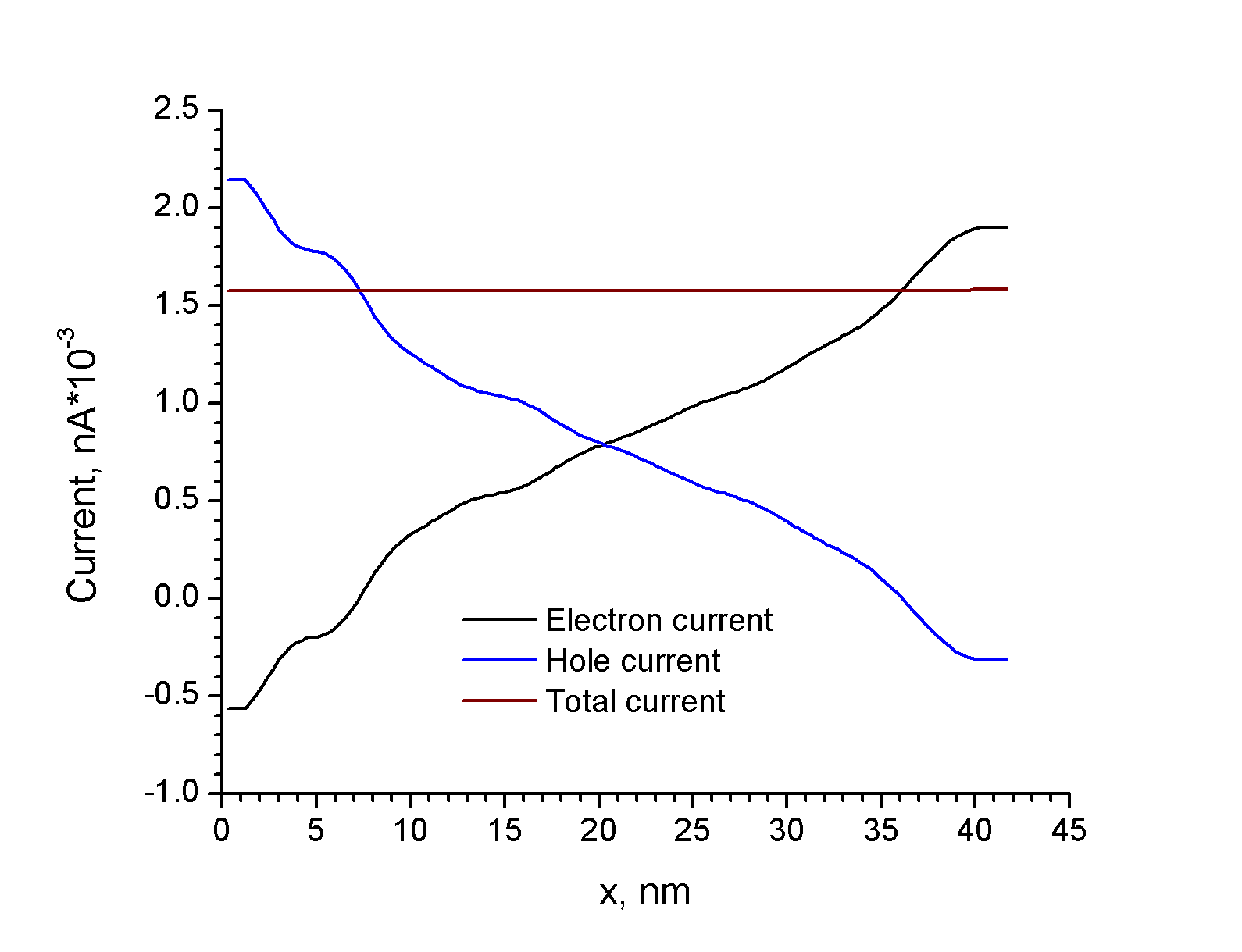}}
  \end{center}
  \caption{Spatial distribution of hole and electron currents at  $~E_\lambda=1.64$ eV,  $~E_\lambda=1.76$ eV, $~E_\lambda=2.00$ eV}
  \label{fig:CBVB}
\end{figure}

\begin{figure}
\centering
\includegraphics[width=300.0px]{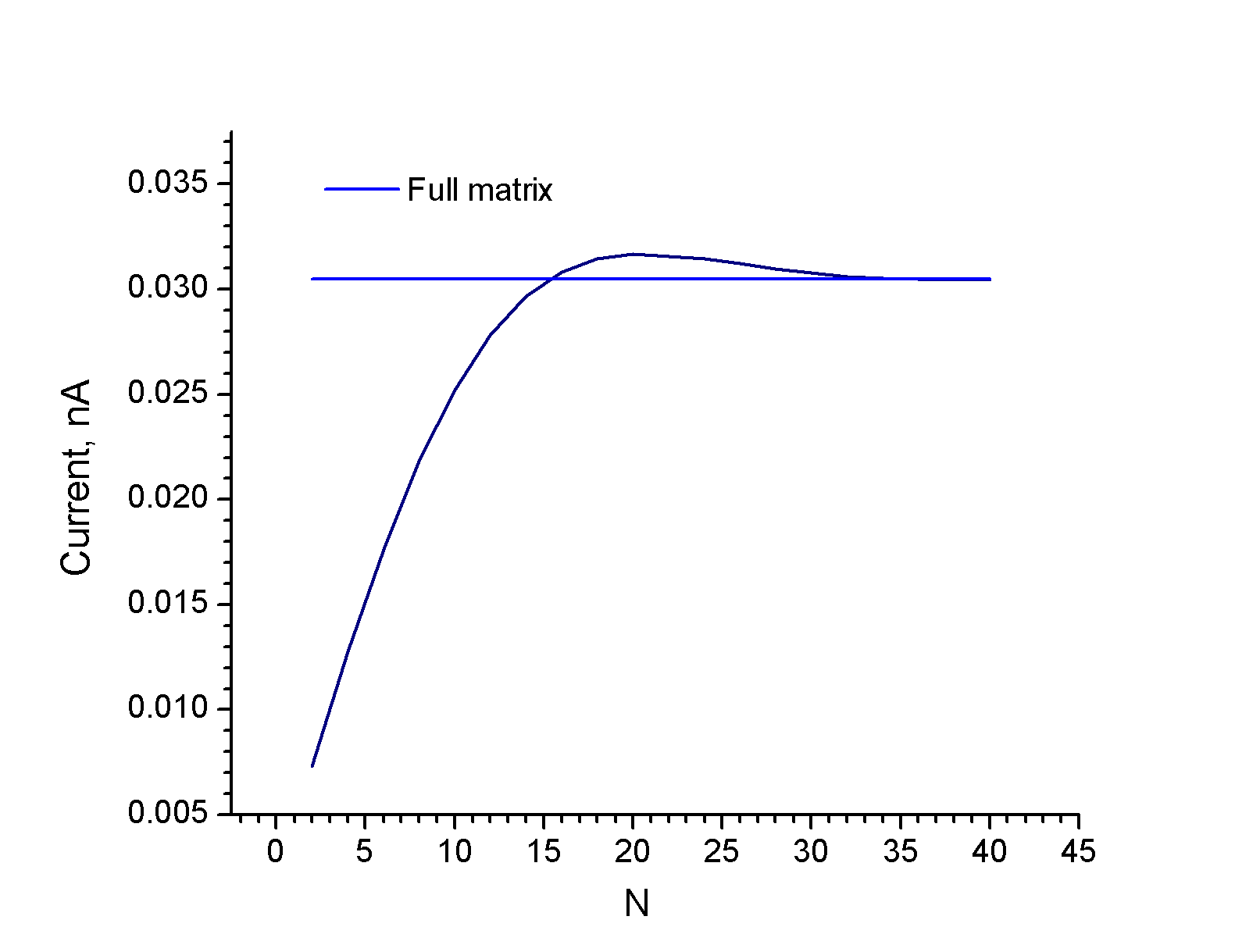}
\caption{Current vs. number of Off-diagonals at $E_{\lambda}=1.5 eV$ at zero bias for p-i-n structure at $I=130$ $W/cm^{2}$ . Blue line corresponds to full matrix $N=139$ }
\label{fig:OFFD}
\end{figure}

\begin{figure}[!]
  \begin{center}
    \subfigure[$N_{d}=8$]{\label{fig:ND-a}\includegraphics[scale=0.5]{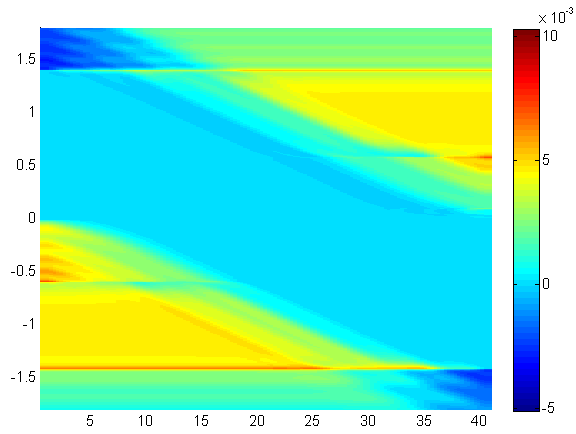}}
    \subfigure[$N_{d}=40$]{\label{fig:ND-b}\includegraphics[scale=0.5]{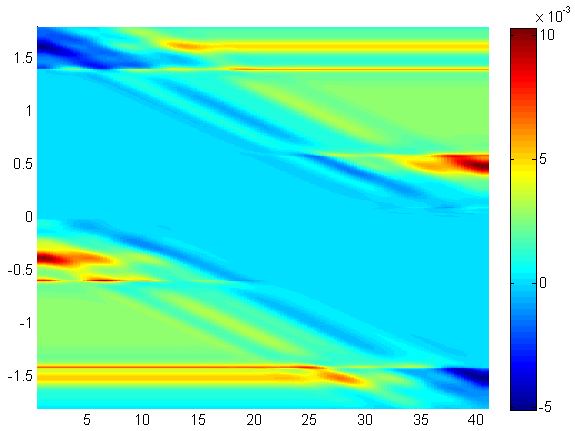}}
  \end{center}
  \caption{Effect of number of off-diagonals on the 2D current. a) $N_{d}=8$ b) $N_{d}=40$, which is approaching full matrix rank}
  \label{fig:ND}
\end{figure}

In addition,  it can be seen that spontaneous emission does not play a role in the device of this length, since typical lifetime of the carrier before spontaneously emitting a photon is of the order of nanoseconds which would require much longer device length to experience it. One should mention that under certain bias conditions, it is possible to have spontaneous emission even for the small-sized devices\cite{SteigerPhD}. For our system to experience it, one would go beyond 1.4eV as we have checked in the range of our  biases, and no contributions have been seen.

The peaks in the photocurrent can be explained through interference, which leads to oscillations in the Joint Density of states(JDOS) along energy coordinate.

Figure \ref{fig:IV} shows the IV-curve of the diode. Dark current in our case is negligible at given biases and the only current computed is the photo-current. Exponential decay can be explained in terms of absorption decrease as bias increases. The order of magnitude and IV curve have a reasonable agreement (taking into account just two conducting modes and device dimensions) with the experimental data on the GaAs pn-diode photo-response\cite{GaAsExp}.

Figure \ref{fig:2Ds} shows 2D spectral currents at different photon energies. One can see that going from $E_{\lambda}$=1.4 eV to $E_{\lambda}$=1.64 eV, the channel current increases with respect to band-edge currents due to increase of available DOS. On the other hand, at $E_{\lambda}$=1.76 eV current in diode  flows near band-edges. This phenomenon can be explained by spatial oscillations of LDOS. The photo-current is mainly due to carrier photo-generation and inter-subbband recombination.
This is more pronounced at Figures \ref{fig:2Ds-e}  and \ref{fig:2Ds-f}. One can see spatial current oscillations in the valence band , which can also be seen  in the energy coordinate. In addition to this, one can see some negative components at contacts. PN Diodes are known for it's rectification properties, which is not the case here and does deteriorate the device performance. The way to bypass this either use potential barriers such as it is done in the work of Henrickson et.al \cite{henrickson} or by using carrier-selective contacts\cite{carrierselective}.

These features are also seen on the Figure \ref{fig:CBVB}. This is mostly pronounced in Fig. \ref{fig:CBVB-b} and Fig.\ref{fig:CBVB-c}. In Fig. \ref{fig:CBVB-b} one  can see oscillations in the hole and electron current density along the length of the device. Figure \ref{fig:CBVB-c} is richer in features such as curvature change over the diode length. These features can be explained by the local generation-recombination rates. Eq.(\ref{eq:RR}) is basically microscopic analog of the macroscopic balance equation\cite{AerbSuper}. Left hand side represents divergence of the particle current(electron) , in our case it is just photo-current due to ballistic photo-extracted  carriers. Right hand side is the energy and volume integrated local generation recombination rates.  It can be rewritten in the form of Eq's. (\ref{eq:RR1}) and (\ref{eq:RR2}). Now physical origin of these terms becomes clear. By looking at r.h.s of the Eq.(\ref{eq:RR2}) one can see two terms. First one corresponds to the total inscattering rate(generation) at that energy, whereas second term gives total ouscattering(recombination) rate at that energy.One should note that by using r.h.s. of the equation (\ref{eq:RR}), and treating particularly first term as  inscattering  and second as outscattering term may lead to nonphysical features in the local recombination-generation spectra such as appearance  of negative inscattering. Although, physical origin of this phenomena is unclear, but we believe it is due to a fact that coherence length in our case is infinite since we are working within ballistic regime. Including non-radiative phase-breaking mechanisms such as phonon-scattering and working in the limit where mean-free path is smaller than device region should remove this problem.
\begin{equation}
\label{eq:RR}
\nabla \cdot \mathbf{J}_{ph}=\frac{1}{V\pi \hbar} \int d^{3}r\int d^{3}r^{\prime }\int dE\{\Sigma_{e-ph}
^{<}(r,r^{\prime },E)G^{>}(r^{\prime },r,E)-\Sigma_{e-ph}^{>}(r,r^{\prime
},E)G^{<}(r^{\prime },r,E)\}
\end{equation}

\begin{equation}
\label{eq:RR1}
\nabla \cdot \mathbf{J}_{ph}=\frac{1}{V\pi \hbar}\int tr[\mathbf{\Sigma}_{e-ph}^{<}(E)\mathbf{G}^{>}(E)-%
\mathbf{\Sigma}_{e-ph}^{>}(E)\mathbf{G}^{<}(E)]dE=\frac{1}{V\pi}\int tr\{\mathbf{I\}}dE
\end{equation}

\begin{equation}
\label{eq:RR2}
\mathbf{I}(E)=\frac{1}{\hbar}tr[\mathbf{\Sigma}_{e-ph}^{<}(E)\mathbf{G}^{>}(E)]-\frac{1}{\hbar}tr[\mathbf{%
\Sigma}_{e-ph}^{>}(E)\mathbf{G}^{<}(E)]
\end{equation}

One can also see that in the case of $E_{\lambda}$=1.76 eV in  Figure \ref{fig:CBVB-b} divergence of the electron and hole current separately is negligible.
This indicates that no photocurrent is being generated in the channel region, which is in agreement with  Figure \ref{fig:2Ds-e}.
By tuning the $E_{\lambda}$ one can make the diode  channel conducting or contact conducting(regions just below valence band and above conduction band). This is also seen in the profiles of the electron and hole currents at different photon energies.

Finally, we also investigate the effect of the non-locality of the $\mathbf{\Sigma }_{e-ph}$ and it'-s effect on the current densities. It is important to note that recursive implementation will fail in this case since it only accounts for the first off-diagonal terms. As can be seen in Fig.\ref{fig:OFFD}, by keeping only 2-off diagonal blocks we have a value of the current which is $2.3\%$   of the total current, which is in agreement with the results of Pourfath et. al.\cite{NonlSelf}. In addition to this, we have also computed 2D distribution of current density. It is seen that spatial current oscillations are lost in case of $N_{d}=8$  as compared to $N_{d}=40$,  where one reconstructs original picture. The reason for this is wave-like \cite{NonlSelf} behaviour of the self-energy which gives phase-coherent response in the limit of the $N_{d} \rightarrow N_{total}$ , where $N_{total}$ is 139 in our case. Phase coherence is lost once just few off-diagonals are retained. Although, some\cite{AerbSuper,Aerb2} works consider that only a portion of the device is being illuminated, in this case it can be shown that self-energies $\mathbf{\Sigma }_{e-ph}$ become only band-diagonal, thus giving possibility of inclusion of smaller number of off-band diagonal blocks.

\section{Conclusions and Outlook.}
We have presented the theoretical framework for the computation the opto-electronic response of the 1D devices in the \textbf{k} $\cdot${} \textbf{p} model, with particular emphasis on the {GaAs} material. Extension of the model from direct bandgap materials to the indirect ones for sub-10 nm 1D systems is straightforward but tedious. Phonon scattering incorporation is straightforward. It is shown that in the phase-coherent limit one observes carrier (e, h) spatial current oscillations. It has also been-shown that local recombination-generation rates may lead to negative components of the current at the leads, which in itself deteriorates the device performance. In addition, it was shown that current can flow in the channel region or near the band edges and not in the channel depending on the incident photon energy. Moreover, local generation-recombination  rates can form different spatial patterns which is reflected in the features of the photocurrent. Moreover, generation-recombination rates may experience nonphysical behaviour such as negative in-scattering, but we believe this is due to a fact that we are working in the ballistic regime. Effect of the non-locality of the self energy is crucial to the computation of the photo-response both quantitatively and qualitatively. Even two subband model reveals non-trivial behaviour of the electronic response upon light illumination. Numerical results are in reasonable agreement with recent experimental data taking into account number of modes and dimensions of the device. The future work includes more realistic implementation by incorporating more 1D subbands. Penetrable boundaries, band-mixing effects, 3D intraband scattering reflected in 1D inter-subband scattering within CB or VB manifolds, going beyond dipole approximation and more general recursive implementation with finite number of off-diagonals is underway.


\begin{thebibliography}{56}%
\makeatletter
\providecommand \@ifxundefined [1]{%
 \@ifx{#1\undefined}
}%
\providecommand \@ifnum [1]{%
 \ifnum #1\expandafter \@firstoftwo
 \else \expandafter \@secondoftwo
 \fi
}%
\providecommand \@ifx [1]{%
 \ifx #1\expandafter \@firstoftwo
 \else \expandafter \@secondoftwo
 \fi
}%
\providecommand \natexlab [1]{#1}%
\providecommand \enquote  [1]{``#1''}%
\providecommand \bibnamefont  [1]{#1}%
\providecommand \bibfnamefont [1]{#1}%
\providecommand \citenamefont [1]{#1}%
\providecommand \href@noop [0]{\@secondoftwo}%
\providecommand \href [0]{\begingroup \@sanitize@url \@href}%
\providecommand \@href[1]{\@@startlink{#1}\@@href}%
\providecommand \@@href[1]{\endgroup#1\@@endlink}%
\providecommand \@sanitize@url [0]{\catcode `\\12\catcode `\$12\catcode
  `\&12\catcode `\#12\catcode `\^12\catcode `\_12\catcode `\%12\relax}%
\providecommand \@@startlink[1]{}%
\providecommand \@@endlink[0]{}%
\providecommand \url  [0]{\begingroup\@sanitize@url \@url }%
\providecommand \@url [1]{\endgroup\@href {#1}{\urlprefix }}%
\providecommand \urlprefix  [0]{URL }%
\providecommand \Eprint [0]{\href }%
\providecommand \doibase [0]{http://dx.doi.org/}%
\providecommand \selectlanguage [0]{\@gobble}%
\providecommand \bibinfo  [0]{\@secondoftwo}%
\providecommand \bibfield  [0]{\@secondoftwo}%
\providecommand \translation [1]{[#1]}%
\providecommand \BibitemOpen [0]{}%
\providecommand \bibitemStop [0]{}%
\providecommand \bibitemNoStop [0]{.\EOS\space}%
\providecommand \EOS [0]{\spacefactor3000\relax}%
\providecommand \BibitemShut  [1]{\csname bibitem#1\endcsname}%
\let\auto@bib@innerbib\@empty
\bibitem [{\citenamefont {J.~Ramanujam}(2011)}]{Ramanujam}%
  \BibitemOpen
  \bibfield  {author} {\bibinfo {author} {\bibfnamefont {A.~V.}\ \bibnamefont
  {J.~Ramanujam}, \bibfnamefont {D.~Shiri}},\ }\href {\doibase
  doi:10.1166/mex.2011.1013} {\bibfield  {journal} {\bibinfo  {journal}
  {Materials Express}\ }\textbf {\bibinfo {volume} {1}},\ \bibinfo {pages}
  {105--126} (\bibinfo {year} {2011})}\BibitemShut {NoStop}%
\bibitem [{\citenamefont {Verma}, \citenamefont {Buin},\ and\ \citenamefont
  {Anantram}(2009)}]{verma}%
  \BibitemOpen
  \bibfield  {author} {\bibinfo {author} {\bibfnamefont {A.}~\bibnamefont
  {Verma}}, \bibinfo {author} {\bibfnamefont {A.~K.}\ \bibnamefont {Buin}}, \
  and\ \bibinfo {author} {\bibfnamefont {M.~P.}\ \bibnamefont {Anantram}},\
  }\href {\doibase 10.1063/1.3264629} {\bibfield  {journal} {\bibinfo
  {journal} {J. Appl. Phys.}\ }\textbf {\bibinfo {volume} {106}},\ \bibinfo
  {eid} {113713} (\bibinfo {year} {2009})}\BibitemShut {NoStop}%
\bibitem [{\citenamefont {Cao}\ \emph {et~al.}(2009)\citenamefont {Cao},
  \citenamefont {White}, \citenamefont {Park}, \citenamefont {Schuller},
  \citenamefont {Clemens},\ and\ \citenamefont {Brongersma}}]{Cao}%
  \BibitemOpen
  \bibfield  {author} {\bibinfo {author} {\bibfnamefont {L.}~\bibnamefont
  {Cao}}, \bibinfo {author} {\bibfnamefont {J.~S.}\ \bibnamefont {White}},
  \bibinfo {author} {\bibfnamefont {J.-S.}\ \bibnamefont {Park}}, \bibinfo
  {author} {\bibfnamefont {J.~A.}\ \bibnamefont {Schuller}}, \bibinfo {author}
  {\bibfnamefont {B.~M.}\ \bibnamefont {Clemens}}, \ and\ \bibinfo {author}
  {\bibfnamefont {M.~L.}\ \bibnamefont {Brongersma}},\ }\href@noop {}
  {\bibfield  {journal} {\bibinfo  {journal} {Nat. Mater.}\ }\textbf {\bibinfo
  {volume} {8}},\ \bibinfo {pages} {643--647} (\bibinfo {year}
  {2009})}\BibitemShut {NoStop}%
\bibitem [{\citenamefont {Agarwal}\ and\ \citenamefont
  {Lieber}(2006)}]{Lieber}%
  \BibitemOpen
  \bibfield  {author} {\bibinfo {author} {\bibfnamefont {R.}~\bibnamefont
  {Agarwal}}\ and\ \bibinfo {author} {\bibfnamefont {C.~M.}\ \bibnamefont
  {Lieber}},\ }\href {\doibase 10.1007/s00339-006-3720-z} {\bibfield  {journal}
  {\bibinfo  {journal} {Appl. Phys. A}\ }\textbf {\bibinfo {volume} {85}},\
  \bibinfo {pages} {209--215} (\bibinfo {year} {2006})}\BibitemShut {NoStop}%
\bibitem [{\citenamefont {Thunich}\ \emph {et~al.}(2009)\citenamefont
  {Thunich}, \citenamefont {Prechtel}, \citenamefont {Spirkoska}, \citenamefont
  {Abstreiter}, \citenamefont {{Fontcuberta i Morral}},\ and\ \citenamefont
  {Holleitner}}]{thunich}%
  \BibitemOpen
  \bibfield  {author} {\bibinfo {author} {\bibfnamefont {S.}~\bibnamefont
  {Thunich}}, \bibinfo {author} {\bibfnamefont {L.}~\bibnamefont {Prechtel}},
  \bibinfo {author} {\bibfnamefont {D.}~\bibnamefont {Spirkoska}}, \bibinfo
  {author} {\bibfnamefont {G.}~\bibnamefont {Abstreiter}}, \bibinfo {author}
  {\bibfnamefont {A.}~\bibnamefont {{Fontcuberta i Morral}}}, \ and\ \bibinfo
  {author} {\bibfnamefont {A.~W.}\ \bibnamefont {Holleitner}},\ }\href
  {\doibase 10.1063/1.3193540} {\bibfield  {journal} {\bibinfo  {journal}
  {Appl. Phys. Lett.}\ }\textbf {\bibinfo {volume} {95}},\ \bibinfo {eid}
  {083111} (\bibinfo {year} {2009})}\BibitemShut {NoStop}%
\bibitem [{\citenamefont {Zardo}\ \emph {et~al.}(2012)\citenamefont {Zardo},
  \citenamefont {Yazji}, \citenamefont {Marini}, \citenamefont {Uccelli},
  \citenamefont {i~Morral}, \citenamefont {Abstreiter},\ and\ \citenamefont
  {Postorino}}]{zardo}%
  \BibitemOpen
  \bibfield  {author} {\bibinfo {author} {\bibfnamefont {I.}~\bibnamefont
  {Zardo}}, \bibinfo {author} {\bibfnamefont {S.}~\bibnamefont {Yazji}},
  \bibinfo {author} {\bibfnamefont {C.}~\bibnamefont {Marini}}, \bibinfo
  {author} {\bibfnamefont {E.}~\bibnamefont {Uccelli}}, \bibinfo {author}
  {\bibfnamefont {A.~F.}\ \bibnamefont {i~Morral}}, \bibinfo {author}
  {\bibfnamefont {G.}~\bibnamefont {Abstreiter}}, \ and\ \bibinfo {author}
  {\bibfnamefont {P.}~\bibnamefont {Postorino}},\ }\href {\doibase
  10.1021/nn300228u} {\bibfield  {journal} {\bibinfo  {journal} {ACS Nano}\
  }\textbf {\bibinfo {volume} {6}},\ \bibinfo {pages} {3284--3291} (\bibinfo
  {year} {2012})}\BibitemShut {NoStop}%
\bibitem [{\citenamefont {Shiri}\ \emph {et~al.}(2012)\citenamefont {Shiri},
  \citenamefont {Verma}, \citenamefont {Selvakumar},\ and\ \citenamefont
  {Anantram}}]{Dar}%
  \BibitemOpen
  \bibfield  {author} {\bibinfo {author} {\bibfnamefont {D.}~\bibnamefont
  {Shiri}}, \bibinfo {author} {\bibfnamefont {A.}~\bibnamefont {Verma}},
  \bibinfo {author} {\bibfnamefont {C.~R.}\ \bibnamefont {Selvakumar}}, \ and\
  \bibinfo {author} {\bibfnamefont {M.~P.}\ \bibnamefont {Anantram}},\
  }\href@noop {} {\bibfield  {journal} {\bibinfo  {journal} {Scientific
  Reports}\ }\textbf {\bibinfo {volume} {2}},\ \bibinfo {pages} {461} (\bibinfo
  {year} {2012})}\BibitemShut {NoStop}%
\bibitem [{\citenamefont {Demichel}\ \emph {et~al.}(2010)\citenamefont
  {Demichel}, \citenamefont {Heiss}, \citenamefont {Bleuse}, \citenamefont
  {Mariette},\ and\ \citenamefont {i~Morral}}]{demichel}%
  \BibitemOpen
  \bibfield  {author} {\bibinfo {author} {\bibfnamefont {O.}~\bibnamefont
  {Demichel}}, \bibinfo {author} {\bibfnamefont {M.}~\bibnamefont {Heiss}},
  \bibinfo {author} {\bibfnamefont {J.}~\bibnamefont {Bleuse}}, \bibinfo
  {author} {\bibfnamefont {H.}~\bibnamefont {Mariette}}, \ and\ \bibinfo
  {author} {\bibfnamefont {A.~F.}\ \bibnamefont {i~Morral}},\ }\href@noop {}
  {\bibfield  {journal} {\bibinfo  {journal} {Appl. Phys. Lett.}\ }\textbf
  {\bibinfo {volume} {97}},\ \bibinfo {eid} {201907} (\bibinfo {year}
  {2010})}\BibitemShut {NoStop}%
\bibitem [{\citenamefont {Chang}\ \emph {et~al.}(2012)\citenamefont {Chang},
  \citenamefont {Chi}, \citenamefont {Yao}, \citenamefont {Huang},
  \citenamefont {Chen}, \citenamefont {Theiss}, \citenamefont {Bushmaker},
  \citenamefont {LaLumondiere}, \citenamefont {Yeh}, \citenamefont {Povinelli},
  \citenamefont {Zhou}, \citenamefont {Dapkus},\ and\ \citenamefont
  {Cronin}}]{Chang}%
  \BibitemOpen
  \bibfield  {author} {\bibinfo {author} {\bibfnamefont {C.~C.}\ \bibnamefont
  {Chang}}, \bibinfo {author} {\bibfnamefont {C.~Y.}\ \bibnamefont {Chi}},
  \bibinfo {author} {\bibfnamefont {M.}~\bibnamefont {Yao}}, \bibinfo {author}
  {\bibfnamefont {N.}~\bibnamefont {Huang}}, \bibinfo {author} {\bibfnamefont
  {C.~C.}\ \bibnamefont {Chen}}, \bibinfo {author} {\bibfnamefont
  {J.}~\bibnamefont {Theiss}}, \bibinfo {author} {\bibfnamefont {A.~W.}\
  \bibnamefont {Bushmaker}}, \bibinfo {author} {\bibfnamefont {S.}~\bibnamefont
  {LaLumondiere}}, \bibinfo {author} {\bibfnamefont {T.~W.}\ \bibnamefont
  {Yeh}}, \bibinfo {author} {\bibfnamefont {M.~L.}\ \bibnamefont {Povinelli}},
  \bibinfo {author} {\bibfnamefont {C.}~\bibnamefont {Zhou}}, \bibinfo {author}
  {\bibfnamefont {P.~D.}\ \bibnamefont {Dapkus}}, \ and\ \bibinfo {author}
  {\bibfnamefont {S.~B.}\ \bibnamefont {Cronin}},\ }\href {\doibase
  10.1021/nl301391h} {\bibfield  {journal} {\bibinfo  {journal} {Nano Lett.}\
  }\textbf {\bibinfo {volume} {12}},\ \bibinfo {pages} {4484--4489} (\bibinfo
  {year} {2012})}\BibitemShut {NoStop}%
\bibitem [{\citenamefont {Sk}\ \emph {et~al.}(2013)\citenamefont {Sk}, ,
  \citenamefont {Ng}, \citenamefont {Huang},\ and\ \citenamefont {Lim}}]{C3CP}%
  \BibitemOpen
  \bibfield  {author} {\bibinfo {author} {\bibfnamefont {M.~A.}\ \bibnamefont
  {Sk}}, , \bibinfo {author} {\bibfnamefont {M.-F.}\ \bibnamefont {Ng}},
  \bibinfo {author} {\bibfnamefont {L.}~\bibnamefont {Huang}}, \ and\ \bibinfo
  {author} {\bibfnamefont {K.~H.}\ \bibnamefont {Lim}},\ }\href {\doibase
  10.1039/C3CP43530J} {\bibfield  {journal} {\bibinfo  {journal} {Phys. Chem.
  Chem. Phys.}\ }\textbf {\bibinfo {volume} {15}},\ \bibinfo {pages}
  {5927--5935} (\bibinfo {year} {2013})}\BibitemShut {NoStop}%
\bibitem [{\citenamefont {Colombo}\ \emph {et~al.}(2009)\citenamefont
  {Colombo}, \citenamefont {Hei$\beta$}, \citenamefont {Gratzel},\ and\
  \citenamefont {i~Morral}}]{colombo}%
  \BibitemOpen
  \bibfield  {author} {\bibinfo {author} {\bibfnamefont {C.}~\bibnamefont
  {Colombo}}, \bibinfo {author} {\bibfnamefont {M.}~\bibnamefont {Hei$\beta$}},
  \bibinfo {author} {\bibfnamefont {M.}~\bibnamefont {Gratzel}}, \ and\
  \bibinfo {author} {\bibfnamefont {A.~F.}\ \bibnamefont {i~Morral}},\ }\href
  {\doibase 10.1063/1.3125435} {\bibfield  {journal} {\bibinfo  {journal}
  {Appl. Phys. Lett.}\ }\textbf {\bibinfo {volume} {94}},\ \bibinfo {eid}
  {173108} (\bibinfo {year} {2009})}\BibitemShut {NoStop}%
\bibitem [{\citenamefont {Witzigmann}\ \emph {et~al.}(2009)\citenamefont
  {Witzigmann}, \citenamefont {Veprek}, \citenamefont {Steiger},\ and\
  \citenamefont {Kupec}}]{AAA}%
  \BibitemOpen
  \bibfield  {author} {\bibinfo {author} {\bibfnamefont {B.}~\bibnamefont
  {Witzigmann}}, \bibinfo {author} {\bibfnamefont {R.~G.}\ \bibnamefont
  {Veprek}}, \bibinfo {author} {\bibfnamefont {S.}~\bibnamefont {Steiger}}, \
  and\ \bibinfo {author} {\bibfnamefont {J.}~\bibnamefont {Kupec}},\ }\href
  {\doibase 10.1007/s10825-009-0274-2} {\bibfield  {journal} {\bibinfo
  {journal} {J. Comput. Electron.}\ }\textbf {\bibinfo {volume} {8}},\ \bibinfo
  {pages} {389--397} (\bibinfo {year} {2009})}\BibitemShut {NoStop}%
\bibitem [{\citenamefont {Fedoseyev}, \citenamefont {Turowski},\ and\
  \citenamefont {Wartak}(2007)}]{Fedoseyev}%
  \BibitemOpen
  \bibfield  {author} {\bibinfo {author} {\bibfnamefont {A.~I.}\ \bibnamefont
  {Fedoseyev}}, \bibinfo {author} {\bibfnamefont {M.}~\bibnamefont {Turowski}},
  \ and\ \bibinfo {author} {\bibfnamefont {M.~S.}\ \bibnamefont {Wartak}},\
  }\href {\doibase doi:10.1166/jno.2007.303} {\bibfield  {journal} {\bibinfo
  {journal} {J. Nanoelectron. Optoe.}\ }\textbf {\bibinfo {volume} {2}},\
  \bibinfo {pages} {234--256} (\bibinfo {year} {2007})}\BibitemShut {NoStop}%
\bibitem [{\citenamefont {Aeberhard}\ and\ \citenamefont {Morf}(2008)}]{Aerb2}%
  \BibitemOpen
  \bibfield  {author} {\bibinfo {author} {\bibfnamefont {U.}~\bibnamefont
  {Aeberhard}}\ and\ \bibinfo {author} {\bibfnamefont {R.~H.}\ \bibnamefont
  {Morf}},\ }\href {\doibase 10.1103/PhysRevB.77.125343} {\bibfield  {journal}
  {\bibinfo  {journal} {Phys. Rev. B}\ }\textbf {\bibinfo {volume} {77}},\
  \bibinfo {pages} {125343} (\bibinfo {year} {2008})}\BibitemShut {NoStop}%
\bibitem [{\citenamefont {Aeberhard}(2011{\natexlab{a}})}]{AerbSuper}%
  \BibitemOpen
  \bibfield  {author} {\bibinfo {author} {\bibfnamefont {U.}~\bibnamefont
  {Aeberhard}},\ }\href@noop {} {\bibfield  {journal} {\bibinfo  {journal}
  {Nanoscale. Res. Lett.}\ }\textbf {\bibinfo {volume} {6}},\ \bibinfo {pages}
  {242} (\bibinfo {year} {2011}{\natexlab{a}})}\BibitemShut {NoStop}%
\bibitem [{\citenamefont {Steiger}(2009)}]{SteigerPhD}%
  \BibitemOpen
  \bibfield  {author} {\bibinfo {author} {\bibfnamefont {S.}~\bibnamefont
  {Steiger}},\ }\emph {\bibinfo {title} {Modelling Nano-LEDs}},\ \href@noop {}
  {\bibinfo {type} {{PhD} dissertation}},\ \bibinfo  {school} {Swiss Federal
  Institute of Technology Zurich (ETHZ)} (\bibinfo {year} {2009})\BibitemShut
  {NoStop}%
\bibitem [{\citenamefont {Henrickson}(2002)}]{henrickson}%
  \BibitemOpen
  \bibfield  {author} {\bibinfo {author} {\bibfnamefont {L.~E.}\ \bibnamefont
  {Henrickson}},\ }\href {\doibase 10.1063/1.1473677} {\bibfield  {journal}
  {\bibinfo  {journal} {J. Appl. Phys.}\ }\textbf {\bibinfo {volume} {91}},\
  \bibinfo {pages} {6273--6281} (\bibinfo {year} {2002})}\BibitemShut {NoStop}%
\bibitem [{\citenamefont {Stewart}\ and\ \citenamefont
  {L\'eonard}(2004)}]{Stewart}%
  \BibitemOpen
  \bibfield  {author} {\bibinfo {author} {\bibfnamefont {D.~A.}\ \bibnamefont
  {Stewart}}\ and\ \bibinfo {author} {\bibfnamefont {F.}~\bibnamefont
  {L\'eonard}},\ }\href {\doibase 10.1103/PhysRevLett.93.107401} {\bibfield
  {journal} {\bibinfo  {journal} {Phys. Rev. Lett.}\ }\textbf {\bibinfo
  {volume} {93}},\ \bibinfo {pages} {107401} (\bibinfo {year}
  {2004})}\BibitemShut {NoStop}%
\bibitem [{\citenamefont {Kane}(1982)}]{Kane82}%
  \BibitemOpen
  \bibfield  {author} {\bibinfo {author} {\bibfnamefont {E.}~\bibnamefont
  {Kane}},\ }\href@noop {} {}edited by\ \bibinfo {editor} {\bibfnamefont
  {W.}~\bibnamefont {Paul}},\ Vol.~\bibinfo {volume} {1}\ (\bibinfo
  {publisher} {North-Holland},\ \bibinfo {address} {Amsterdam},\ \bibinfo
  {year} {1982})\ pp.\ \bibinfo {pages} {193--217}\BibitemShut {NoStop}%
\bibitem [{\citenamefont {Boujdaria}\ \emph {et~al.}(2002)\citenamefont
  {Boujdaria}, \citenamefont {Ridene}, \citenamefont {Radhia}, \citenamefont
  {Zitouni}, \citenamefont {Bouchriha},\ and\ \citenamefont {Fishman}}]{kp714}%
  \BibitemOpen
  \bibfield  {author} {\bibinfo {author} {\bibfnamefont {K.}~\bibnamefont
  {Boujdaria}}, \bibinfo {author} {\bibfnamefont {S.}~\bibnamefont {Ridene}},
  \bibinfo {author} {\bibfnamefont {S.~B.}\ \bibnamefont {Radhia}}, \bibinfo
  {author} {\bibfnamefont {O.}~\bibnamefont {Zitouni}}, \bibinfo {author}
  {\bibfnamefont {H.}~\bibnamefont {Bouchriha}}, \ and\ \bibinfo {author}
  {\bibfnamefont {G.}~\bibnamefont {Fishman}},\ }\href@noop {} {\bibfield
  {journal} {\bibinfo  {journal} {J. Appl. Phys}\ }\textbf {\bibinfo {volume}
  {92}},\ \bibinfo {pages} {2586--2592} (\bibinfo {year} {2002})}\BibitemShut
  {NoStop}%
\bibitem [{\citenamefont {El~kurdi}\ \emph {et~al.}(2003)\citenamefont
  {El~kurdi}, \citenamefont {Fishman}, \citenamefont {Sauvage},\ and\
  \citenamefont {Boucaud}}]{KpSi}%
  \BibitemOpen
  \bibfield  {author} {\bibinfo {author} {\bibfnamefont {M.}~\bibnamefont
  {El~kurdi}}, \bibinfo {author} {\bibfnamefont {G.}~\bibnamefont {Fishman}},
  \bibinfo {author} {\bibfnamefont {S.}~\bibnamefont {Sauvage}}, \ and\
  \bibinfo {author} {\bibfnamefont {P.}~\bibnamefont {Boucaud}},\ }\href
  {\doibase 10.1103/PhysRevB.68.165333} {\bibfield  {journal} {\bibinfo
  {journal} {Phys. Rev. B}\ }\textbf {\bibinfo {volume} {68}},\ \bibinfo
  {pages} {165333} (\bibinfo {year} {2003})}\BibitemShut {NoStop}%
\bibitem [{\citenamefont {Ridene}\ \emph {et~al.}(2001)\citenamefont {Ridene},
  \citenamefont {Boujdaria}, \citenamefont {Bouchriha},\ and\ \citenamefont
  {Fishman}}]{kpSi1}%
  \BibitemOpen
  \bibfield  {author} {\bibinfo {author} {\bibfnamefont {S.}~\bibnamefont
  {Ridene}}, \bibinfo {author} {\bibfnamefont {K.}~\bibnamefont {Boujdaria}},
  \bibinfo {author} {\bibfnamefont {H.}~\bibnamefont {Bouchriha}}, \ and\
  \bibinfo {author} {\bibfnamefont {G.}~\bibnamefont {Fishman}},\ }\href
  {\doibase 10.1103/PhysRevB.64.085329} {\bibfield  {journal} {\bibinfo
  {journal} {Phys. Rev. B}\ }\textbf {\bibinfo {volume} {64}},\ \bibinfo
  {pages} {085329} (\bibinfo {year} {2001})}\BibitemShut {NoStop}%
\bibitem [{\citenamefont {Boujdaria}\ and\ \citenamefont
  {Zitouni}(2004)}]{kpSi30}%
  \BibitemOpen
  \bibfield  {author} {\bibinfo {author} {\bibfnamefont {K.}~\bibnamefont
  {Boujdaria}}\ and\ \bibinfo {author} {\bibfnamefont {O.}~\bibnamefont
  {Zitouni}},\ }\href {\doibase 10.1016/j.ssc.2003.07.011} {\bibfield
  {journal} {\bibinfo  {journal} {Solid State Commun.}\ }\textbf {\bibinfo
  {volume} {129}},\ \bibinfo {pages} {205 -- 210} (\bibinfo {year}
  {2004})}\BibitemShut {NoStop}%
\bibitem [{\citenamefont {Zitouni}, \citenamefont {Boujdaria},\ and\
  \citenamefont {Bouchriha}(2005)}]{kp24}%
  \BibitemOpen
  \bibfield  {author} {\bibinfo {author} {\bibfnamefont {O.}~\bibnamefont
  {Zitouni}}, \bibinfo {author} {\bibfnamefont {K.}~\bibnamefont {Boujdaria}},
  \ and\ \bibinfo {author} {\bibfnamefont {H.}~\bibnamefont {Bouchriha}},\
  }\href@noop {} {\bibfield  {journal} {\bibinfo  {journal} {Semicond. Sci.
  Technol.}\ }\textbf {\bibinfo {volume} {20}},\ \bibinfo {pages} {908}
  (\bibinfo {year} {2005})}\BibitemShut {NoStop}%
\bibitem [{\citenamefont {Cardona}\ and\ \citenamefont
  {Pollak}(1966)}]{Cardona}%
  \BibitemOpen
  \bibfield  {author} {\bibinfo {author} {\bibfnamefont {M.}~\bibnamefont
  {Cardona}}\ and\ \bibinfo {author} {\bibfnamefont {F.~H.}\ \bibnamefont
  {Pollak}},\ }\href {\doibase 10.1103/PhysRev.142.530} {\bibfield  {journal}
  {\bibinfo  {journal} {Phys. Rev.}\ }\textbf {\bibinfo {volume} {142}},\
  \bibinfo {pages} {530--543} (\bibinfo {year} {1966})}\BibitemShut {NoStop}%
\bibitem [{\citenamefont {Bahder}(1990)}]{kp8}%
  \BibitemOpen
  \bibfield  {author} {\bibinfo {author} {\bibfnamefont {T.~B.}\ \bibnamefont
  {Bahder}},\ }\href {\doibase 10.1103/PhysRevB.41.11992} {\bibfield  {journal}
  {\bibinfo  {journal} {Phys. Rev. B}\ }\textbf {\bibinfo {volume} {41}},\
  \bibinfo {pages} {11992--12001} (\bibinfo {year} {1990})}\BibitemShut
  {NoStop}%
\bibitem [{\citenamefont {Foreman}(1997)}]{Foreman}%
  \BibitemOpen
  \bibfield  {author} {\bibinfo {author} {\bibfnamefont {B.~A.}\ \bibnamefont
  {Foreman}},\ }\href@noop {} {\bibfield  {journal} {\bibinfo  {journal} {Phys.
  Rev. B}\ }\textbf {\bibinfo {volume} {56}},\ \bibinfo {pages}
  {R12748--R12751} (\bibinfo {year} {1997})}\BibitemShut {NoStop}%
\bibitem [{\citenamefont {Pidgeon}\ and\ \citenamefont
  {Brown}(1966)}]{ModifLuttinger}%
  \BibitemOpen
  \bibfield  {author} {\bibinfo {author} {\bibfnamefont {C.~R.}\ \bibnamefont
  {Pidgeon}}\ and\ \bibinfo {author} {\bibfnamefont {R.~N.}\ \bibnamefont
  {Brown}},\ }\href {\doibase 10.1103/PhysRev.146.575} {\bibfield  {journal}
  {\bibinfo  {journal} {Phys. Rev.}\ }\textbf {\bibinfo {volume} {146}},\
  \bibinfo {pages} {575--583} (\bibinfo {year} {1966})}\BibitemShut {NoStop}%
\bibitem [{\citenamefont {Shin}(2009)}]{shinMain}%
  \BibitemOpen
  \bibfield  {author} {\bibinfo {author} {\bibfnamefont {M.}~\bibnamefont
  {Shin}},\ }\href@noop {} {\bibfield  {journal} {\bibinfo  {journal} {J. Appl.
  Phys}\ }\textbf {\bibinfo {volume} {106}},\ \bibinfo {eid} {054505} (\bibinfo
  {year} {2009})}\BibitemShut {NoStop}%
\bibitem [{\citenamefont {Pourfath}, \citenamefont {Baumgartner},\ and\
  \citenamefont {Kosina}(2008)}]{NonlSelf}%
  \BibitemOpen
  \bibfield  {author} {\bibinfo {author} {\bibfnamefont {M.}~\bibnamefont
  {Pourfath}}, \bibinfo {author} {\bibfnamefont {O.}~\bibnamefont
  {Baumgartner}}, \ and\ \bibinfo {author} {\bibfnamefont {H.}~\bibnamefont
  {Kosina}},\ }\bibfield  {title} {\enquote {\bibinfo {title} {On the
  non-locality of the electron-photon self-energy: Application to carbon
  nanotube photo-detectors},}\ }in\ \href {\doibase 10.1109/NUSOD.2008.4668261}
  {\emph {\bibinfo {booktitle} {Numerical Simulation of Optoelectronic Devices,
  2008. NUSOD '08. International Conference on}}}\ (\bibinfo {year} {2008})\
  pp.\ \bibinfo {pages} {99--100}\BibitemShut {NoStop}%
\bibitem [{\citenamefont {Lake}\ and\ \citenamefont {Datta}(1992)}]{contacts}%
  \BibitemOpen
  \bibfield  {author} {\bibinfo {author} {\bibfnamefont {R.}~\bibnamefont
  {Lake}}\ and\ \bibinfo {author} {\bibfnamefont {S.}~\bibnamefont {Datta}},\
  }\href {\doibase 10.1103/PhysRevB.45.6670} {\bibfield  {journal} {\bibinfo
  {journal} {Phys. Rev. B}\ }\textbf {\bibinfo {volume} {45}},\ \bibinfo
  {pages} {6670--6685} (\bibinfo {year} {1992})}\BibitemShut {NoStop}%
\bibitem [{\citenamefont {Thompson}, \citenamefont {Rasmussen},\ and\
  \citenamefont {Lookman}(2004)}]{Anderson}%
  \BibitemOpen
  \bibfield  {author} {\bibinfo {author} {\bibfnamefont {R.~B.}\ \bibnamefont
  {Thompson}}, \bibinfo {author} {\bibfnamefont {K.~O.~.}\ \bibnamefont
  {Rasmussen}}, \ and\ \bibinfo {author} {\bibfnamefont {T.}~\bibnamefont
  {Lookman}},\ }\href {\doibase 10.1063/1.1629673} {\bibfield  {journal}
  {\bibinfo  {journal} {J. Chem. Phys}\ }\textbf {\bibinfo {volume} {120}},\
  \bibinfo {pages} {31--34} (\bibinfo {year} {2004})}\BibitemShut {NoStop}%
\bibitem [{\citenamefont {Lake}\ and\ \citenamefont {Pandey}(2007)}]{LakeEn}%
  \BibitemOpen
  \bibfield  {author} {\bibinfo {author} {\bibfnamefont {R.~K.}\ \bibnamefont
  {Lake}}\ and\ \bibinfo {author} {\bibfnamefont {R.~R.}\ \bibnamefont
  {Pandey}},\ }\href@noop {} {\bibfield  {journal} {\bibinfo  {journal}
  {arXiv:cond-mat/0607219v1}\ } (\bibinfo {year} {2007})}\BibitemShut {NoStop}%
\bibitem [{\citenamefont {Mahan}(1987)}]{MahanRev}%
  \BibitemOpen
  \bibfield  {author} {\bibinfo {author} {\bibfnamefont {G.~D.}\ \bibnamefont
  {Mahan}},\ }\href {\doibase 10.1016/0370-1573(87)90004-4} {\bibfield
  {journal} {\bibinfo  {journal} {Phys. Rep.}\ }\textbf {\bibinfo {volume}
  {145}},\ \bibinfo {pages} {251 -- 318} (\bibinfo {year} {1987})}\BibitemShut
  {NoStop}%
\bibitem [{\citenamefont {R.~K.~Lake}\ and\ \citenamefont
  {Jovanovic}(1997)}]{Lake}%
  \BibitemOpen
  \bibfield  {author} {\bibinfo {author} {\bibfnamefont {R.~C.~B.}\
  \bibnamefont {R.~K.~Lake}, \bibfnamefont {G.~Klimeck}}\ and\ \bibinfo
  {author} {\bibfnamefont {D.}~\bibnamefont {Jovanovic}},\ }\href {\doibase
  10.1063/1.365394} {\bibfield  {journal} {\bibinfo  {journal} {J. Appl.
  Phys.}\ }\textbf {\bibinfo {volume} {81}},\ \bibinfo {pages} {7845} (\bibinfo
  {year} {1997})}\BibitemShut {NoStop}%
\bibitem [{\citenamefont {Jiang}, \citenamefont {Wang},\ and\ \citenamefont
  {Li}(2011)}]{paper3}%
  \BibitemOpen
  \bibfield  {author} {\bibinfo {author} {\bibfnamefont {J.-W.}\ \bibnamefont
  {Jiang}}, \bibinfo {author} {\bibfnamefont {J.-S.}\ \bibnamefont {Wang}}, \
  and\ \bibinfo {author} {\bibfnamefont {B.}~\bibnamefont {Li}},\ }\href
  {\doibase 10.1063/1.3531573} {\bibfield  {journal} {\bibinfo  {journal} {J.
  Appl. Phys.}\ }\textbf {\bibinfo {volume} {109}},\ \bibinfo {eid} {014326}
  (\bibinfo {year} {2011})}\BibitemShut {NoStop}%
\bibitem [{\citenamefont {Jiang}\ and\ \citenamefont {Wang}(2011)}]{paper2}%
  \BibitemOpen
  \bibfield  {author} {\bibinfo {author} {\bibfnamefont {J.-W.}\ \bibnamefont
  {Jiang}}\ and\ \bibinfo {author} {\bibfnamefont {J.-S.}\ \bibnamefont
  {Wang}},\ }\href {\doibase 10.1063/1.3671069} {\bibfield  {journal} {\bibinfo
   {journal} {J. Appl. Phys.}\ }\textbf {\bibinfo {volume} {110}},\ \bibinfo
  {eid} {124319} (\bibinfo {year} {2011})}\BibitemShut {NoStop}%
\bibitem [{\citenamefont {Luisier}, \citenamefont {Schenk},\ and\ \citenamefont
  {Fichtner}(2006)}]{luisierMode}%
  \BibitemOpen
  \bibfield  {author} {\bibinfo {author} {\bibfnamefont {M.}~\bibnamefont
  {Luisier}}, \bibinfo {author} {\bibfnamefont {A.}~\bibnamefont {Schenk}}, \
  and\ \bibinfo {author} {\bibfnamefont {W.}~\bibnamefont {Fichtner}},\ }\href
  {\doibase 10.1063/1.2244522} {\bibfield  {journal} {\bibinfo  {journal} {J.
  Appl. Phys.}\ }\textbf {\bibinfo {volume} {100}},\ \bibinfo {eid} {043713}
  (\bibinfo {year} {2006})}\BibitemShut {NoStop}%
\bibitem [{\citenamefont {J.Wang}, \citenamefont {Polizzi},\ and\ \citenamefont
  {Lundstrom}(2004)}]{modeSpace}%
  \BibitemOpen
  \bibfield  {author} {\bibinfo {author} {\bibnamefont {J.Wang}}, \bibinfo
  {author} {\bibfnamefont {E.}~\bibnamefont {Polizzi}}, \ and\ \bibinfo
  {author} {\bibfnamefont {M.}~\bibnamefont {Lundstrom}},\ }\href {\doibase
  10.1063/1.1769089} {\bibfield  {journal} {\bibinfo  {journal} {J. Appl.
  Phys.}\ }\textbf {\bibinfo {volume} {96}},\ \bibinfo {pages} {2192--2203}
  (\bibinfo {year} {2004})}\BibitemShut {NoStop}%
\bibitem [{\citenamefont {Jiang}, \citenamefont {Wang},\ and\ \citenamefont
  {Li}(2009)}]{paper1}%
  \BibitemOpen
  \bibfield  {author} {\bibinfo {author} {\bibfnamefont {J.-W.}\ \bibnamefont
  {Jiang}}, \bibinfo {author} {\bibfnamefont {J.-S.}\ \bibnamefont {Wang}}, \
  and\ \bibinfo {author} {\bibfnamefont {B.}~\bibnamefont {Li}},\ }\href
  {\doibase 10.1103/PhysRevB.80.205429} {\bibfield  {journal} {\bibinfo
  {journal} {Phys. Rev. B}\ }\textbf {\bibinfo {volume} {80}},\ \bibinfo
  {pages} {205429} (\bibinfo {year} {2009})}\BibitemShut {NoStop}%
\bibitem [{\citenamefont {Birner}\ \emph {et~al.}(2007)\citenamefont {Birner},
  \citenamefont {Zibold}, \citenamefont {Andlauer}, \citenamefont {Kubis},
  \citenamefont {Sabathil}, \citenamefont {Trellakis},\ and\ \citenamefont
  {Vogl}}]{nextnano}%
  \BibitemOpen
  \bibfield  {author} {\bibinfo {author} {\bibfnamefont {S.}~\bibnamefont
  {Birner}}, \bibinfo {author} {\bibfnamefont {T.}~\bibnamefont {Zibold}},
  \bibinfo {author} {\bibfnamefont {T.}~\bibnamefont {Andlauer}}, \bibinfo
  {author} {\bibfnamefont {T.}~\bibnamefont {Kubis}}, \bibinfo {author}
  {\bibfnamefont {M.}~\bibnamefont {Sabathil}}, \bibinfo {author}
  {\bibfnamefont {A.}~\bibnamefont {Trellakis}}, \ and\ \bibinfo {author}
  {\bibfnamefont {P.}~\bibnamefont {Vogl}},\ }\href {\doibase
  10.1109/TED.2007.902871} {\bibfield  {journal} {\bibinfo  {journal} {IEEE
  Trans. Electron Devices}\ }\textbf {\bibinfo {volume} {54}},\ \bibinfo
  {pages} {2137--2142} (\bibinfo {year} {2007})}\BibitemShut {NoStop}%
\bibitem [{\citenamefont {Rideau}\ \emph {et~al.}(2009)\citenamefont {Rideau},
  \citenamefont {Feraille}, \citenamefont {Michaillat}, \citenamefont {Niquet},
  \citenamefont {Tavernier},\ and\ \citenamefont
  {Jaouen}}]{EnvelopeFirstBrill}%
  \BibitemOpen
  \bibfield  {author} {\bibinfo {author} {\bibfnamefont {D.}~\bibnamefont
  {Rideau}}, \bibinfo {author} {\bibfnamefont {M.}~\bibnamefont {Feraille}},
  \bibinfo {author} {\bibfnamefont {M.}~\bibnamefont {Michaillat}}, \bibinfo
  {author} {\bibfnamefont {Y.}~\bibnamefont {Niquet}}, \bibinfo {author}
  {\bibfnamefont {C.}~\bibnamefont {Tavernier}}, \ and\ \bibinfo {author}
  {\bibfnamefont {H.}~\bibnamefont {Jaouen}},\ }\href {\doibase
  10.1016/j.sse.2008.08.006} {\bibfield  {journal} {\bibinfo  {journal} {Solid
  State Electron.}\ }\textbf {\bibinfo {volume} {53}},\ \bibinfo {pages} {452
  -- 461} (\bibinfo {year} {2009})}\BibitemShut {NoStop}%
\bibitem [{\citenamefont {Foreman}(1995)}]{Foreman95}%
  \BibitemOpen
  \bibfield  {author} {\bibinfo {author} {\bibfnamefont {B.~A.}\ \bibnamefont
  {Foreman}},\ }\href {\doibase 10.1103/PhysRevB.52.12241} {\bibfield
  {journal} {\bibinfo  {journal} {Phys. Rev. B}\ }\textbf {\bibinfo {volume}
  {52}},\ \bibinfo {pages} {12241--12259} (\bibinfo {year} {1995})}\BibitemShut
  {NoStop}%
\bibitem [{\citenamefont {Burt}(1992)}]{Burt}%
  \BibitemOpen
  \bibfield  {author} {\bibinfo {author} {\bibfnamefont {M.~G.}\ \bibnamefont
  {Burt}},\ }\href@noop {} {\bibfield  {journal} {\bibinfo  {journal} {J.
  Phys.: Condens.Matter}\ }\textbf {\bibinfo {volume} {4}},\ \bibinfo {pages}
  {6651} (\bibinfo {year} {1992})}\BibitemShut {NoStop}%
\bibitem [{\citenamefont {Yang}\ and\ \citenamefont {Chang}(2005)}]{CutoffK}%
  \BibitemOpen
  \bibfield  {author} {\bibinfo {author} {\bibfnamefont {W.}~\bibnamefont
  {Yang}}\ and\ \bibinfo {author} {\bibfnamefont {K.}~\bibnamefont {Chang}},\
  }\href {\doibase 10.1103/PhysRevB.72.233309} {\bibfield  {journal} {\bibinfo
  {journal} {Phys. Rev. B}\ }\textbf {\bibinfo {volume} {72}},\ \bibinfo
  {pages} {233309} (\bibinfo {year} {2005})}\BibitemShut {NoStop}%
\bibitem [{\citenamefont {Foreman}(2007)}]{Foreman75}%
  \BibitemOpen
  \bibfield  {author} {\bibinfo {author} {\bibfnamefont {B.~A.}\ \bibnamefont
  {Foreman}},\ }\href {\doibase 10.1103/PhysRevB.75.235331} {\bibfield
  {journal} {\bibinfo  {journal} {Phys. Rev. B}\ }\textbf {\bibinfo {volume}
  {75}},\ \bibinfo {pages} {235331} (\bibinfo {year} {2007})}\BibitemShut
  {NoStop}%
\bibitem [{\citenamefont {Veprek}, \citenamefont {Steiger},\ and\ \citenamefont
  {Witzigmann}(2007)}]{Parameters}%
  \BibitemOpen
  \bibfield  {author} {\bibinfo {author} {\bibfnamefont {R.~G.}\ \bibnamefont
  {Veprek}}, \bibinfo {author} {\bibfnamefont {S.}~\bibnamefont {Steiger}}, \
  and\ \bibinfo {author} {\bibfnamefont {B.}~\bibnamefont {Witzigmann}},\
  }\href {\doibase 10.1103/PhysRevB.76.165320} {\bibfield  {journal} {\bibinfo
  {journal} {Phys. Rev. B}\ }\textbf {\bibinfo {volume} {76}},\ \bibinfo
  {pages} {165320} (\bibinfo {year} {2007})}\BibitemShut {NoStop}%
\bibitem [{\citenamefont {Kolokolov}, \citenamefont {Li},\ and\ \citenamefont
  {Ning}(2003)}]{Kolokolov}%
  \BibitemOpen
  \bibfield  {author} {\bibinfo {author} {\bibfnamefont {K.~I.}\ \bibnamefont
  {Kolokolov}}, \bibinfo {author} {\bibfnamefont {J.}~\bibnamefont {Li}}, \
  and\ \bibinfo {author} {\bibfnamefont {C.~Z.}\ \bibnamefont {Ning}},\ }\href
  {\doibase 10.1103/PhysRevB.68.161308} {\bibfield  {journal} {\bibinfo
  {journal} {Phys. Rev. B}\ }\textbf {\bibinfo {volume} {68}},\ \bibinfo
  {pages} {161308} (\bibinfo {year} {2003})}\BibitemShut {NoStop}%
\bibitem [{\citenamefont {Cartoix\`{a}}, \citenamefont {Ting},\ and\
  \citenamefont {McGill}(2003)}]{real_spurious_meshing}%
  \BibitemOpen
  \bibfield  {author} {\bibinfo {author} {\bibfnamefont {X.}~\bibnamefont
  {Cartoix\`{a}}}, \bibinfo {author} {\bibfnamefont {D.~Z.-Y.}\ \bibnamefont
  {Ting}}, \ and\ \bibinfo {author} {\bibfnamefont {T.~C.}\ \bibnamefont
  {McGill}},\ }\href {\doibase 10.1063/1.1555833} {\bibfield  {journal}
  {\bibinfo  {journal} {J. Appl. Phys.}\ }\textbf {\bibinfo {volume} {93}},\
  \bibinfo {pages} {3974--3981} (\bibinfo {year} {2003})}\BibitemShut {NoStop}%
\bibitem [{\citenamefont {Eissfeller}\ and\ \citenamefont {Vogl}(2011)}]{FEM}%
  \BibitemOpen
  \bibfield  {author} {\bibinfo {author} {\bibfnamefont {T.}~\bibnamefont
  {Eissfeller}}\ and\ \bibinfo {author} {\bibfnamefont {P.}~\bibnamefont
  {Vogl}},\ }\href {\doibase 10.1103/PhysRevB.84.195122} {\bibfield  {journal}
  {\bibinfo  {journal} {Phys. Rev. B}\ }\textbf {\bibinfo {volume} {84}},\
  \bibinfo {pages} {195122} (\bibinfo {year} {2011})}\BibitemShut {NoStop}%
\bibitem [{\citenamefont {Guo}, \citenamefont {Alam},\ and\ \citenamefont
  {Yoon}(2006)}]{Nanotube_Light}%
  \BibitemOpen
  \bibfield  {author} {\bibinfo {author} {\bibfnamefont {J.}~\bibnamefont
  {Guo}}, \bibinfo {author} {\bibfnamefont {M.~A.}\ \bibnamefont {Alam}}, \
  and\ \bibinfo {author} {\bibfnamefont {Y.}~\bibnamefont {Yoon}},\ }\href
  {\doibase 10.1063/1.2189827} {\bibfield  {journal} {\bibinfo  {journal}
  {Appl. Phys. Lett.}\ }\textbf {\bibinfo {volume} {88}},\ \bibinfo {eid}
  {133111} (\bibinfo {year} {2006})}\BibitemShut {NoStop}%
\bibitem [{\citenamefont {Aeberhard}(2011{\natexlab{b}})}]{AerbCoh1}%
  \BibitemOpen
  \bibfield  {author} {\bibinfo {author} {\bibfnamefont {U.}~\bibnamefont
  {Aeberhard}},\ }\href {\doibase 10.1186/1556-276X-6-242} {\bibfield
  {journal} {\bibinfo  {journal} {Nanoscale Res. Lett.}\ }\textbf {\bibinfo
  {volume} {6}},\ \bibinfo {pages} {242} (\bibinfo {year}
  {2011}{\natexlab{b}})}\BibitemShut {NoStop}%
\bibitem [{\citenamefont {Pereira}\ and\ \citenamefont
  {Henneberger}(1996)}]{HohCoh}%
  \BibitemOpen
  \bibfield  {author} {\bibinfo {author} {\bibfnamefont {M.~F.}\ \bibnamefont
  {Pereira}}\ and\ \bibinfo {author} {\bibfnamefont {K.}~\bibnamefont
  {Henneberger}},\ }\href {\doibase 10.1103/PhysRevB.53.16485} {\bibfield
  {journal} {\bibinfo  {journal} {Phys. Rev. B}\ }\textbf {\bibinfo {volume}
  {53}},\ \bibinfo {pages} {16485--16496} (\bibinfo {year} {1996})}\BibitemShut
  {NoStop}%
\bibitem [{\citenamefont {Aeberhard}(2011{\natexlab{c}})}]{AerbCoh}%
  \BibitemOpen
  \bibfield  {author} {\bibinfo {author} {\bibfnamefont {U.}~\bibnamefont
  {Aeberhard}},\ }\href {\doibase 10.1007/s10825-011-0375-6} {\bibfield
  {journal} {\bibinfo  {journal} {Journal of Computational Electronics}\
  }\textbf {\bibinfo {volume} {10}},\ \bibinfo {pages} {394--413} (\bibinfo
  {year} {2011}{\natexlab{c}})}\BibitemShut {NoStop}%
\bibitem [{\citenamefont {Lysov}\ \emph {et~al.}(2011)\citenamefont {Lysov},
  \citenamefont {Gutsche}, \citenamefont {Offer}, \citenamefont {Regolin},
  \citenamefont {Prost},\ and\ \citenamefont {Tegude}}]{GaAsExp}%
  \BibitemOpen
  \bibfield  {author} {\bibinfo {author} {\bibfnamefont {A.}~\bibnamefont
  {Lysov}}, \bibinfo {author} {\bibfnamefont {C.}~\bibnamefont {Gutsche}},
  \bibinfo {author} {\bibfnamefont {M.}~\bibnamefont {Offer}}, \bibinfo
  {author} {\bibfnamefont {I.}~\bibnamefont {Regolin}}, \bibinfo {author}
  {\bibfnamefont {W.}~\bibnamefont {Prost}}, \ and\ \bibinfo {author}
  {\bibfnamefont {F.~J.~.}\ \bibnamefont {Tegude}},\ }\bibfield  {title}
  {\enquote {\bibinfo {title} {The optoelectronic performance of axial and
  radial gaas nanowire pn-diodes},}\ }in\ \href@noop {} {\emph {\bibinfo
  {booktitle} {Compound Semiconductor Week (CSW/IPRM), 2011 and 23rd
  International Conference on Indium Phosphide and Related Materials}}}\
  (\bibinfo {year} {2011})\ pp.\ \bibinfo {pages} {1--3}\BibitemShut {NoStop}%
\bibitem [{\citenamefont {Aeberhard}(2012)}]{carrierselective}%
  \BibitemOpen
  \bibfield  {author} {\bibinfo {author} {\bibfnamefont {U.}~\bibnamefont
  {Aeberhard}},\ }\href@noop {} {\bibfield  {journal} {\bibinfo  {journal}
  {Optical and Quantum Electronics}\ }\textbf {\bibinfo {volume} {44}},\
  \bibinfo {pages} {133--140} (\bibinfo {year} {2012})}\BibitemShut {NoStop}%
\end{thebibliography}
\end{document}